\documentclass{aa}

\usepackage[T1]{fontenc}
\usepackage{ae,aecompl}
\usepackage{xtab}
\usepackage{multirow}
\usepackage{times}
\usepackage{upgreek}
\usepackage{graphicx}
\usepackage{epstopdf}
\usepackage{amsmath}
\usepackage{amssymb}
\usepackage[section]{placeins}
\usepackage{color}
\usepackage{gensymb}
\usepackage{longtable}
\usepackage{natbib}

\newcommand{\uHz}{$\upmu$Hz}
\newcommand{\teff}{T$_{\rm eff}$}
\newcommand{\logg}{$\log$\,g}
\newcommand{\loggcms}{$\log${(g/\cmss)}}

\newcommand{\gaia}{{\it Gaia}}
\newcommand{\tess}{TESS}

\newcommand{\cmss}{\mbox{$\mbox{cm}\,\mbox{s}^{-2}$}}    % centimetres/second^2

\newcommand{\logy}{$\log$(n(He)/n(H))}

\begin{document} 

\title{Short-period pulsating hot-subdwarf stars observed by TESS}
\subtitle{II. Northern ecliptic hemisphere}

\author{A.S.\,Baran\inst{1,2}, S.\,Charpinet\inst{3}, R.H.\,{\O}stensen\inst{2,4}, M.D.\,Reed\inst{2}, V.\,Van Grootel\inst{5}, C.\,Lyu\inst{3}, J.H.\,Telting\inst{6,7}, P.\,N\'emeth\inst{8,9}} 
%D.\,Kilkenny\inst{13}

%\email{andysbaran@gmail.com}
\institute {Astronomical Observatory, University of Warsaw, Al. Ujazdowskie 4, 00-478 Warszawa, Poland
\and ARDASTELLA Research Collaboration, Missouri State University, Springfield, MO\,65897, USA
\and IRAP, CNRS, UPS, CNES, Université de Toulouse, 14 av. Edouard Belin, 31400 Toulouse, France
\and Recogito AS, Storgaten 72, N-8200 Fauske, Norway
\and Space sciences, Technologies and Astrophysics Research (STAR) Institute, Université de Liège, 19C Allée du 6 Août, B-4000 Liège, Belgium
\and Nordic Optical Telescope, Rambla Jos{\'e} Ana Fern{\'a}ndez P{\'e}rez 7, 38711 Bre{\~n}a Baja, Spain
\and Department of Physics and Astronomy, Aarhus University, NyMunkegade 120, DK-8000 Aarhus C, Denmark
\and Astronomical Institute of the Czech Academy of Sciences, Fri\v{c}ova 298, CZ-251\,65 Ond\v{r}ejov, Czech Republic
\and Astroserver.org, F\H{o} t\'er 1, 8533 Malomsok, Hungary
}
\date{}

% \abstract{}{}{}{}{}
% 5 {} token are mandatory

\abstract
{We present results of a continuation of our Transiting Exoplanet Survey Satellite (\tess) search for short-period pulsations in compact stellar objects observed during Years 2 and 4 of the \tess\ mission that targeted the northern ecliptic hemisphere. For many of the targets, we exploit unpublished spectroscopic data to confirm or determine the object's spectral classification. From the \tess\ photometry, we identify 50 short-period hot-subdwarf pulsators, including 35 sdB and 15 sdOB stars. The sample contains 26 pulsators not known before the \tess\ mission. Nine stars show signals at both low and high frequencies, and are therefore ``hybrid'' pulsators. For each pulsator, we report the list of prewhitened frequencies and we show amplitude spectra calculated from the \tess\ data. We attempt to identify possible multiplets caused by stellar rotation, and we report five candidates with rotation periods between 11 and 46\,d. Having the search for p-mode pulsating hot subdwarfs in \tess\ Sectors\,1\,--\,60 done, we discuss the completeness of the study, as well as instability strip and the evolutionary status of the stars we found. We also compare the distribution of pulsation periods as a function of effective temperature and surface gravity with theoretical predictions. We find that the percentage of undetected pulsators in the \tess\ mission increases with decreasing brightnesses of stars, reaching 25\% near the 15$^{\rm th}$ magnitude. When comparing the distribution of hot subdwarfs in the \logg-\teff\ plane with stellar models, we underline the importance of a proper treatment of the hydrogen-rich envelope composition (strongly affected by microscopic diffusion processes). We also emphasize that the stellar mass is a significant factor in understanding the instability strip. The $p$-mode instability strip is confirmed to be narrower than predicted by non-adiabatic calculations based on models incorporating equilibrium between gravitational settling and radiative levitation for iron. This implies that competing mixing processes ignored in these models must play a role to reduce the amount of levitating iron in the stellar envelope. Interestingly, we find that the coolest p-mode pulsators with \teff\ $\lesssim$ 30,000\,K (including the hybrid ones) tend to cluster around the Terminal Age of the Extreme Horizontal Branch of canonical mass ($\sim 0.47$ $M_\odot$), a trend expected from the non-adiabatic pulsation calculations. Otherwise, the overall pulsation period distributions tend to reproduce the predicted trends in \teff\ and \logg.}


\authorrunning{A.S.\,Baran et al.}
\titlerunning{P-mode hot-subdwarf pulsators in the northern ecliptic hemisphere}
\maketitle

\section{Introduction}
\label{intro}
\citet[][hereafter: Paper\,I]{baran23} reported their detection of 43 hot-subdwarf pressure (p) mode pulsators in the southern ecliptic hemisphere observed with the Transiting Exoplanet Survey Satellite (\tess). The sample includes 32 subdwarf B (sdB) stars, eight subdwarf OB (sdOB) stars, two subdwarf O (sdO) stars, and, significantly, one He-sdOB star, which is the first of this kind to show short-period pulsations. An introduction to all types of hot subdwarfs are provided in Paper\,I.

We continued our effort to search for hot-subdwarf p-mode pulsators in the northern ecliptic hemisphere observed by \tess. Our ultimate goal is to report the census of the p-mode-dominated hot-subdwarf stars surveyed by the \tess\ satellite, along with an ensemble analysis on the presence of p-mode pulsators among hot subdwarfs.

\section{Classification spectra}
\label{sec:spectra}
To verify or establish the spectral classification for some of the objects, new spectra were obtained using the ALFOSC spectrograph on the 2.56\,m Nordic Optical Telescope (NOT). We mainly used grism\,\#18 that samples the Balmer series from H$_\beta$ to shorter wavelengths, spanning 345\,--\,535\,nm, and we used a 1-arcsec slit resulting in R\,=\,1000 spectral resolution. These spectra also serve to obtain estimates of the effective temperature, surface gravity and surface He/H, using the classical LTE model atmospheres of \citet{heber00}. The obtained values are quoted in the context of the individual stars.

\section{\tess\ photometry}
\label{data}
We follow the same approach as explained in Paper\,I. Basically, we downloaded all available data of our targets from the Barbara A. Mikulski Archive for Space Telescopes (MAST)\footnote{archive.stsci.edu}. Our preference is to use the USC (ultra-short cadence of 20\,s integration) data, if available, with which the entire p-mode frequency range is sampled. In the case of the SC (short cadence of 120\,s) data, the Nyquist frequency (at 4167\uHz) is in the middle of the p-mode region and aliasing becomes a serious issue. We used PDCSAP\_FLUX, which is corrected for on-board systematics and neighbors' contributions to the overall flux. We clipped fluxes at 4.5$\upsigma$ to remove outliers, de-trended long-term variations (on the order of days) with spline fitting, and calculated the amplitude of the flux variations using the relation A[ppt]=1000*({\rm flux}/{<{\rm flux}>}-1), where ppt is parts-per-thousand.

\section{Fourier analysis}
\label{fourier}
We used the same prewhitening techniques with a detection threshold as described in Paper\,I. Based on results presented by \cite{baran21}, we adopted a detection threshold at signal-to-noise (S/N) of 4.5 times the median noise level to both the SC and USC data sets, regardless of the data coverage. Nevertheless, frequencies with S/N\,$\leq$\,5.5 (corresponding to a FAP\,=\,0.1\%) should be considered tentative. To search for significant signal in the amplitude spectra, we used the FELIX package \citep{charpinet10,zong16} and other dedicated scripts. We followed the same procedure as explained in Paper\,I. At first, an automatic search was performed, looking for significant variations beyond 1\,500 \uHz where $p$-modes are usually found. Then, an individual check was carried out to establish if detected variations are indeed consistent with $p$-mode pulsations. We detected signals in the p-mode region in 50 hot subdwarfs. Twelve targets were observed only in the SC mode, while 38 were observed in the USC mode. We found 11 new detections in SC data and 16 new detections in USC data, while the remaining one\,(SC) and 22\,(USC) hot-subdwarf pulsators were already known prior to the \tess\ mission. We present the list of all p-mode hot subdwarf pulsators in Table\,\ref{tab:targets_all}. This table also provides the spectral type (SpT), either from the literature or from new spectroscopy that we acquired recently. The classification convention we applied is explained in Paper\,I, basically we marked those stars with detectable \ion{He}{I}, or no \ion{He}{I} lines at all, with a B class. Those stars with detectable \ion{He}{II} at 4686\,\AA\ with class OB, and those stars with strong \ion{He}{II} and no \ion{He}{I} with class O. Below we describe each target, including previously-published pulsation properties, followed by amplitude spectra and the list of frequencies detected from the \tess\ light curves.

\begin{table}
\centering
\caption{The list of pulsating hot subdwarfs found in the \tess\ data. No USC data were available prior to Sector\,27, hence data in sectors marked in parentheses were not used in our analysis. New pulsators are marked with \textbf{bold} font.
%Asterisks denote that the USC data were also collected, however only the SC data show signals.
}
\label{tab:targets_all}
\begin{tabular}{|c|r|r|c|}
\hline
&\multicolumn{1}{c|}{TIC} & \multicolumn{1}{c|}{Sector} & \multicolumn{1}{c|}{SpT} \\
\hline
\multirow{12}{*}{\rotatebox[origin=c]{90}{SC targets}}
& \textbf{968226}    & 42 & sdB \\
& \textbf{16993518}  & 51 & sdOB \\
& \textbf{85145647}  & 20,47 & sdB \\
& \textbf{97286494}  & 44\,--\,47 & sdB \\
& \textbf{154818961} & 57 & sdB \\
& \textbf{157141133} & 54 & sdB \\
&          186484490 & 17 & sdB \\
& \textbf{199715319} & 40,49\,--\,53,56,57,59 & sdB \\
%& 219509849 & 25,50$^*$\,--\,52$^*$ & sdB \\
& \textbf{222892604} & 54 & sdB \\
& \textbf{248776104} & 55 & sdB \\
%& 311521201 & 25,26,52,53 & sdB \\ source of the peaks?
& \textbf{331656308} & 52,53,59 & sdB \\
& \textbf{392092589} & 44\,--\,46 & sdB \\
\hline
\multirow{38}{*}{\rotatebox[origin=c]{90}{USC targets}}
& \textbf{4632676}   & 49 & sdOB \\
&           26291471 & 51 & sdOB \\
& \textbf{55753808}  & (21),(22),41,48 & sdOB \\
&           56863037 & (21),47 & sdB \\
&           60985176 & (17),42,43 & sdOB \\
%&  63126689 & 40,41,54 & sdOB \\
& \textbf{63168679}  & 44,46 & sdOB \\
&           68495594 & (20),44,45,60 & sdOB \\
&           82359147 & (24)\,--\,(26),51,52 & sdB \\
& \textbf{88484868}  & (20),47,60 & sdB \\
&           88565376 & (20),47,60 & sdB \\
&           90960668 & 56 & sdB \\
& \textbf{114196505} & 54 & sdB \\
%& 117354072 & 44\,--\,46 & sdB \\ long periods listed in Table
&          136975077 & (15),(16),55,56 & sdB \\
& \textbf{137502282} & (20),40,47,53,60 & sdOB \\
&          138618727 & (14),(15),(21),41,48 & sdB \\
& \textbf{142398823} & 41,48,49 & sdB \\
&          159644241 & (14),(15),40,41,54,55 & sdB \\
&          165312944 & (15),(22),48,49 & sdB \\
& \textbf{166054500} & (22),(23),48,49 & sdB \\
&          175402069 & (23),46,50 & sdB \\
& \textbf{178081355} & 58 & sdOB \\
&          191442416 & (17),57 & sdB \\
& \multirow{2}{*}{\textbf{202354658}} & (15),(16),(18),(22)\,--\,(24), & \multirow{2}{*}{sdB} \\
&                    & 48\,--\,51,58 & \\
&          207440586 & (16),(22)\,--\,(25),49\,--\,52,56 & sdOB \\
&          219492314 & (24),(25),51,52 & sdOB \\
& \textbf{240868270} & (17),(18),58 & sdOB \\
&          266013993 & 42,43 & sdB \\
&          273255412 & 54 & sdB \\
&          284692897 & 40,41,54 & sdB \\
& \textbf{309807601} & 49,50 & sdB \\
&          310937915 & 42 & sdOB \\
&          355754830 & 60 & sdB \\
& \textbf{357232133} & 47,60 & sdOB \\
& \textbf{364966239} & 41,47,54--57,60 & sdOB \\
&          397595169 & 56 & sdB \\
& \multirow{2}{*}{424720852} & (14)\,--\,(16),(20),(23), & \multirow{2}{*}{sdB} \\
&                    & 40,41,50,54\,--\,57,60 & \\
& \multirow{2}{*}{\textbf{441725813}} & (14)\,--\,(25), & \multirow{2}{*}{sdB} \\
&                    & 40,41,47\,--\,52,55\,--\,60 & \\
& \multirow{2}{*}{471015194} & (14),(16),(17),(20),(23), & \multirow{2}{*}{sdB} \\
&                    & 40,41,47,50,54\,--\,57,60 & \\
\hline
\end{tabular}
\end{table}

\subsection{Targets observed in the SC mode}

\begin{figure*}
\centering
\includegraphics[width=\hsize]{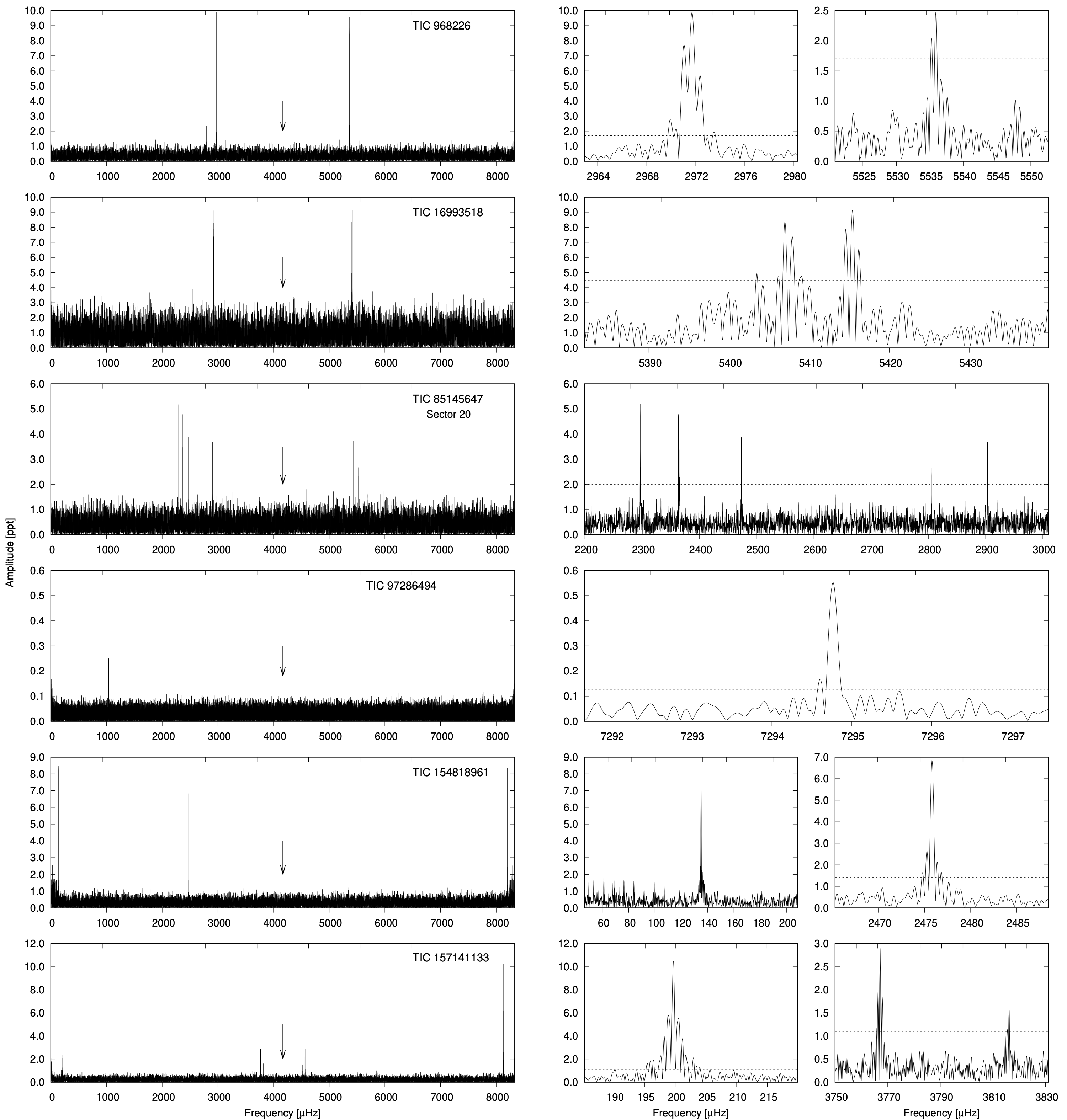}
\caption{Amplitude spectra of the targets observed in the SC mode. If sector is not specified, we included all data. {\it Left panels:} Frequency range, up to twice the Nyquist frequency. The arrows point at the Nyquist frequency of 4166.67\,\uHz. {\it Right panels:} Close-ups of the detected frequencies.}
\label{fig:SC_ft1}
\end{figure*}

For stars with SC mode data, the Nyquist frequency is at 4167\,\uHz. As in Paper\,I, we decided to prewhiten frequencies that are found in the super-Nyquist region. We tried to discern sub- from super-Nyquist, by first relying on previously-published results for known pulsators and then using a combination of amplitude, multi-sector
observations (if available), and peak shape in the Fourier transform (FT). We show amplitude spectra of SC-observed pulsating subdwarf B stars (sdBV) in Figure\,\ref{fig:SC_ft1} and \ref{fig:SC_ft2} and provide the seismic properties of the detected pulsation frequencies, $f_1$ to $f_N$ in Table\,\ref{tab:SCfreq}, as well as any detected binary orbital signal as $\Omega$ and its harmonic $2\Omega$.

\paragraph{TIC\,968226}
(PHL\,211) is a new sdB pulsator. The spectral classification was revealed by \citet{kilkenny84}, who marked it as an sdB star based on Str{\"o}mgren indices. Our fit to a spectrum taken with the NOT
in 2022 gives \teff\,=\,28\,982(198)\,K, \loggcms\,=\,5.42(3) and \logy\,=\,-3.15(9), and we classified the star as an sdB. \tess\ observed the star during Sector\,42. We detected two frequencies in the p-mode region provided in Table\,\ref{tab:SCfreq} and shown in Fig.\,\ref{fig:SC_ft1}.

\paragraph{TIC\,16993518}
(FBS\,1539+355) is a new sdOB pulsator. \citet{perez16} found the star as a hot subdwarf candidate with \teff\,=\,32\,500\,K. Later, based on photometry, \citet{geier17} classified the star as an sdB. Our fit to a spectrum taken with the NOT in 2022 gives \teff\,=\,32\,194(187)\,K, \loggcms\,=\,5.37(4) and \logy\,=\,-2.27(9), and we classified the star as an sdOB. \tess\ observed the star during Sector\,51. We detected three frequencies in the p-mode region shown in Fig.\,\ref{fig:SC_ft1}, with frequencies in Table\,\ref{tab:SCfreq}.

\begin{table}
\centering
\caption{List of frequencies detected in the targets observed only with the SC. If sector is not specified, we included all data.}
\label{tab:SCfreq}
\begin{tabular}{cllrr}
\hline\hline
\multirow{2}{*}{ID} & \multicolumn{1}{c}{Frequency} & \multicolumn{1}{c}{Period} & \multicolumn{1}{c}{Amp.} & \multirow{2}{*}{S/N}\\
& \multicolumn{1}{c}{[\uHz]} & \multicolumn{1}{c}{[sec]} & \multicolumn{1}{c}{[ppt]} & \\
\hline\hline
\multicolumn{5}{c}{{\bf TIC\,968226}}\\
f$_{\rm 1}$ & 2971.735(8)  &    336.5038(9)  &   9.89(32) &  26.4\\
f$_{\rm 2}$ & 5535.884(30) &    180.6396(10) &   2.51(32) &   6.7\\
\hline
\multicolumn{5}{c}{{\bf TIC\,16993518}}\\
f$_{\rm 1}$ & 5406.991(25) &    184.9458(9)  &   8.8(9)   &   8.8\\
f$_{\rm 2}$ & 5409.845(48) &    184.8482(16) &   4.6(9)   &   4.6\\
f$_{\rm 3}$ & 5415.405(24) &    184.6584(8)  &   9.1(9)   &   9.1\\
\hline
\multicolumn{5}{c}{{\bf TIC\,85145647\,--\,Sector\,20}}\\
f$_{\rm 1}$ & 2296.635(17) &    435.4197(33) &   5.19(38) &  11.7\\
f$_{\rm 2}$ & 2363.520(20) &    423.0977(36) &   4.61(38) &  10.4\\
f$_{\rm 3}$ & 2364.626(30) &    422.900(5)   &   3.26(38) &   7.3\\
f$_{\rm 4}$ & 2365.62(6)   &    422.723(10)  &   1.61(38) &   3.6\\
f$_{\rm 5}$ & 2473.358(23) &    404.3086(38) &   3.87(38) &   8.7\\
f$_{\rm 6}$ & 2805.127(34) &    356.4900(43) &   2.64(38) &   6.0\\
f$_{\rm 7}$ & 2903.264(24) &    344.4399(29) &   3.69(38) &   8.3\\
\hline
\multicolumn{5}{c}{{\bf TIC\,85145647\,--\,Sector\,47}}\\
f$_{\rm 1}$ & 2296.617(20) &    435.4231(39) &   6.0(5)   &   9.7\\
f$_{\rm 2}$ & 2363.493(14) &    423.1026(25) &   8.6(5)   &  14.0\\
\hline
\multicolumn{5}{c}{{\bf TIC\,97286494}}\\
f$_{\rm 1}$ & 7294.7690(26) & 137.084532(48) &  0.552(24) &  19.7\\
\hline
\multicolumn{5}{c}{{\bf TIC\,154818961}}\\
$\Omega$  &  67.5236(34) &  14809.6(7)  &   1.51(27) &   4.8\\
2$\Omega$ &  135.0472    &   7404.8     &   8.51(27) &  26.9\\
f$_{\rm 1}$    & 2475.812(8)  & 403.9079(14) &   6.84(27) &  21.6\\
\hline
\multicolumn{5}{c}{{\bf TIC\,157141133}}\\
f$_{\rm 1}$ & 199.6893(44) &     5007.78(11) &  10.49(21) &  43.4\\
f$_{\rm 2}$ & 3767.119(16) &    265.4549(11) &   2.90(21) &  12.0\\
f$_{\rm 3}$ & 3816.152(28) &    262.0441(19) &   1.63(21) &   6.7\\
\hline
\multicolumn{5}{c}{{\bf TIC\,186484490}}\\
$\Omega$  &  45.9351(43) &    21769.9(2.0) &  56.3(1) &  47.1\\
2$\Omega$ &  91.8702     &    10884.95     &   8.6(1) &   7.2\\
f$_{\rm 1}$    & 2965.506(26) &    337.2105(29) &   9.7(1) &   8.1\\
\hline
\multicolumn{5}{c}{{\bf TIC\,199715319\,--\,Sector\,49\,--\,53,56,57,59}}\\
f$_{\rm 1}$ & 2809.4228(16) &    355.94500(21) &   3.76(30) &  10.8\\
\hline
\multicolumn{5}{c}{{\bf TIC\,222892604}}\\
f$_{\rm 1}$ & 4968.6121(35) &  201.26345(14) &  10.03(11) & 104.5\\
f$_{\rm 2}$ & 4969.116(6)   &  201.24306(23) &   6.36(11) &  66.2\\
f$_{\rm 3}$ & 4979.745(30)  &  200.8135(12)  &   0.62(8)  &   6.4\\
f$_{\rm 4}$ & 5038.476(13)  &  198.4727(5)   &   1.47(8)  &  15.3\\
f$_{\rm 5}$ & 5223.378(40)  &  191.4470(15)  &   0.47(8)  &   4.9\\
\hline
\multicolumn{5}{c}{{\bf TIC\,248776104}}\\
f$_{\rm 1}$ & 3735.552(26) &    267.6980(19) &   5.4(6) &   7.3\\
f$_{\rm 2}$ & 3838.819(42) &    260.4968(29) &   3.3(6) &   4.5\\
\hline
\multicolumn{5}{c}{{\bf TIC\,331656308\,--\,Sector\,52\,--\,53}}\\
f$_{\rm 1}$ &  109.235(12)  &    9154.5(1.0)  &   2.52(25) &   8.4\\
f$_{\rm 2}$ & 2672.733(16)  &    374.1489(23) &   1.91(25) &   6.4\\
f$_{\rm 3}$ & 2810.7229(41) &    355.7804(5)  &   7.44(25) &  24.9\\
f$_{\rm 4}$ & 2874.6505(20) &   347.86838(24) &  15.43(25) &  51.7\\
\hline
\multicolumn{5}{c}{{\bf TIC\,331656308\,--\,Sector\,59}}\\
f$_{\rm 1}$ &  147.351(40) &  6786.5(1.8)  &   1.69(27) &   5.3\\
f$_{\rm 2}$ &  278.870(32) &  3585.91(41)  &   2.09(27) &   6.5\\
f$_{\rm 3}$ &  325.350(35) &  3073.61(33)  &   1.94(27) &   6.1\\
f$_{\rm 4}$ & 2810.714(8)  &  355.7815(10) &   8.33(27) &  26.1\\
f$_{\rm 5}$ & 2874.478(10) &  347.8893(12) &   6.62(27) &  20.8\\
\hline
\multicolumn{5}{c}{{\bf TIC\,392092589}}\\
f$_{\rm 1}$ &  214.340(13) &   4665.49(27) &   1.42(22) &   5.4\\
f$_{\rm 2}$ &  284.692(9)  &   3512.57(11) &   2.00(22) &   7.6\\
f$_{\rm 3}$ &  308.856(11) &   3237.75(11) &   1.65(22) &   6.3\\
f$_{\rm 4}$ & 2669.558(7)  &  374.5938(10) &   2.42(22) &   9.2\\
\hline\hline
\end{tabular}
\end{table}

\paragraph{TIC\,85145647}
(FBS\,0725+623) is a new sdB pulsator. Based on photometry, \citet{geier17} classified the star as an sdB. Our fit to a spectrum taken with the INT
%\footnote{All INT spectra used in this work were obtained with ??, Grism\,\#??, and a ?? arcsec slit, giving R\,=\,?? and spanning ??\,nm.}
in 2014 gives \teff\,=\,32\,287(656)\,K, \loggcms\,=\,5.41(12) and \logy\,=\,-3.03(39), indicating an sdB classification. \tess\ observed the star during Sectors\,20 and 47. We detected seven frequencies in Sector\,20 and two frequencies in Sector\,47, which differed in amplitude. We show the amplitude spectrum calculated from Sector\,20 data in Fig.\,\ref{fig:SC_ft1}, and we list the prewhitened frequencies in data from both Sectors in Table\,\ref{tab:SCfreq}. Three frequencies f$_2$, f$_3$ and f$_4$, make a slightly asymmetric triplet. The signal-to-noise (S/N) ratio of f$_4$ is only 3.6, which may not be an intrinsic signal. The splittings on both sides are 1.11 and 0.99\,\uHz\ and assuming the splitting is caused by rotation, the average rotation period is 11.0(7)\,days.

\paragraph{TIC\,97286494}
(GALEX\,J081110.8+273420) is a new sdB pulsator. Based on photometry, \citet{geier17} classified the star as an sdB+F. \tess\ observed the star during Sectors 44\,--\,47 from which we detected only one frequency in the p-mode region. We show the amplitude spectrum in Fig.\,\ref{fig:SC_ft1}, and list the prewhitened frequency in Table\,\ref{tab:SCfreq}.

\paragraph{TIC\,154818961}
(\gaia\,DR3 1935962732084366592) is a new sdB pulsator. Our fit to a spectrum taken with the NOT in 2023 gives \teff\,=\,28\,330(281)\,K, \loggcms\,=\,5.29(4) and \logy\,=\,-1.71(5), and we classified the star as an sdB. \tess\ observed the star during Sector\,57. We detected three frequencies and based on amplitudes and profiles we selected those in the sub-Nyquist region. We interpret the two low frequencies to be a binary frequency and its first harmonic. We show the amplitude spectrum in Fig.\,\ref{fig:SC_ft1}, and list the prewhitened frequencies in Table\,\ref{tab:SCfreq}. The middle panel shows an additional comb of low amplitude frequencies; we found that TIC\,2041859698 (\gaia\,DR3 1935962732084366464) is the source of those frequencies, which we interpret to be caused by eclipses. We estimate the orbital period of this eclipsing binary system to be near 1.5\,d.

\paragraph{TIC\,157141133}
(\gaia\,DR3 6904512841490841472) is a new sdB pulsator. Our fit to a spectrum taken with the ING in 2014 gives \teff\,=\,31\,076(207)\,K, \loggcms\,=\,5.60(4) and \logy\,=\,-2.97(16), and we classified the star as an sdB. \tess\ observed the star during Sector\,54. We detected three frequencies in the sub-Nyquist region. Frequency f$_1$ is likely a binary signature. We show the amplitude spectrum in Fig.\,\ref{fig:SC_ft1}, and the prewhitened frequencies in Table\,\ref{tab:SCfreq}.

\paragraph{TIC\,186484490}
(FBS\,0117+396) is a known sdB pulsator. It was originally identified as sdB-O in the First Byurakan Sky Survey \citep{abrahamian90}. \cite{ostensen13} derived atmospheric parameters and classified the star as an sdB. The same authors reported a discovery of the star to be an sdB+dM binary system with a pulsating primary. Two high and eight low frequency signals were reported and the orbital period was estimated to be 0.252013(13)\,days. The binarity was confirmed with radial velocity data. \tess\ observed the star during Sector\,17. We detected a binary frequency and its first harmonic. The orbital period we measured is 0.251966(23)\,days and we detected only one high frequency signal, frequency $f_1$ reported by \citet{ostensen13}, and in the sub-Nyquist region. We show the amplitude spectrum in Fig.\,\ref{fig:SC_ft2}, and list the prewhitened frequencies in Table\,\ref{tab:SCfreq}.

\paragraph{TIC\,199715319}
(PG\,1656+553) is a new sdB pulsator. The star was classified as an sdB star in the Palomar-Green (PG) survey \citep{green86} and later confirmed by \citet{kilkenny88}. Our fit to a spectrum taken with the NOT in 2023 gives \teff\,=\,28\,144(310)\,K, \loggcms\,=\,5.32(4) and \logy\,=\,-3.15(10), and we classified the star as an sdB. \tess\ observed the star during Sectors\,40,49\,--\,53,56\,--\,57 and 59. We detect only one frequency, listed in Table\,\ref{tab:SCfreq} and show the amplitude spectrum calculated from all sectors except 40 in Fig.\,\ref{fig:SC_ft2}. 

\paragraph{TIC\,222892604}
(LS\, IV+06$^\circ$\,5) is a new sdB pulsator. \citet{nassau63} classified it as a main-sequence  B star. We obtained a NOT spectrum in 2022 and re-classify it as an sdB star. Our fit gives \teff\,=\,30\,363(188)\,K, \loggcms\,=\,5.47(3) and \logy\,=\,-3.08(13). \tess\ observed the star during Sector\,54. We detected five frequencies with two of them being close to each other, possibly part of a multiplet. The frequency splitting is 0.5039\,\uHz, giving a rotation period of either 23 or 46\,days (for $\Delta m$ of 1 or 2, respectively). We show the amplitude spectrum in Fig.\,\ref{fig:SC_ft2}, and list the prewhitened frequencies in Table\,\ref{tab:SCfreq}.

\paragraph{TIC\,248776104}
(PG\,2052$-$003) is a new sdB pulsator. The star was classified as an sdB star in the Palomar-Green (PG) survey \citep{green86}. Our fit to a spectrum taken with the NOT in 2022 gives \teff\,=\,32\,891(205)\,K, \loggcms\,=\,5.52(4) and \logy\,=\,-2.89(14), and we classified the star as an sdB. \tess\ observed the star during Sector\,55. We detected two frequencies, which we list in Table\,\ref{tab:SCfreq} and show the amplitude spectrum in Fig.\,\ref{fig:SC_ft2}. 

\paragraph{TIC\,331656308}
(FBS\,0433+759) is a new sdB pulsator. Our fit to a spectrum taken with the NOT in 2022 gives \teff\,=\,28\,411(230)\,K, \loggcms\,=\,5.36(3) and \logy\,=\,-2.95(11), and we classified the star as an sdB. \tess\ observed the star during Sectors\,52, 53 and 59. The frequency content significantly varies over time and we decided to analyze each contiguous set separately. Therefore, Sectors\,52 and 53 were merged but Sector\,59 data were treated separately. We detected four and five frequencies in the former and latter data sets, respectively. One/three frequencies were detected in the g-mode region, which makes the star a hybrid pulsator. The total power of p-mode signals during Sector\,59 is smaller than during Sectors\,52\,--\,53, while the power of g-mode signals are quite opposite. Energy transferred between modes through nonlinear coupling is a possible explanation for such a variable pulsation spectrum \citep{zong16}. We show the amplitude spectrum for Sector\,59 data in Fig.\,\ref{fig:SC_ft2}, and we separately list the prewhitened frequencies from both data sets in Table\,\ref{tab:SCfreq}.

\paragraph{TIC\,392092589}
(Ton\,4) is a new sdB pulsator. It was classified as a hot subdwarf by \citet{perez16}. Our fit to a spectrum taken with the NOT in 2015 gives \teff\,=\,27\,654(280)\,K, \loggcms\,=\,5.40(4) and \logy\,=\,-2.97(10), and we confirm the sdB classification. \tess\ observed the star during Sectors\,44\,--\,46. We detected four frequencies, three g-modes and one p-mode. Two additional g-mode frequencies may be seen in the middle panel, however their amplitudes do not meet our detection limit. We show the amplitude spectrum in Fig.\,\ref{fig:SC_ft2}, and list the prewhitened frequencies in Table\,\ref{tab:SCfreq}.

\subsection{Targets observed in the USC mode}
\begin{figure*}
\centering
\includegraphics[width=\hsize]{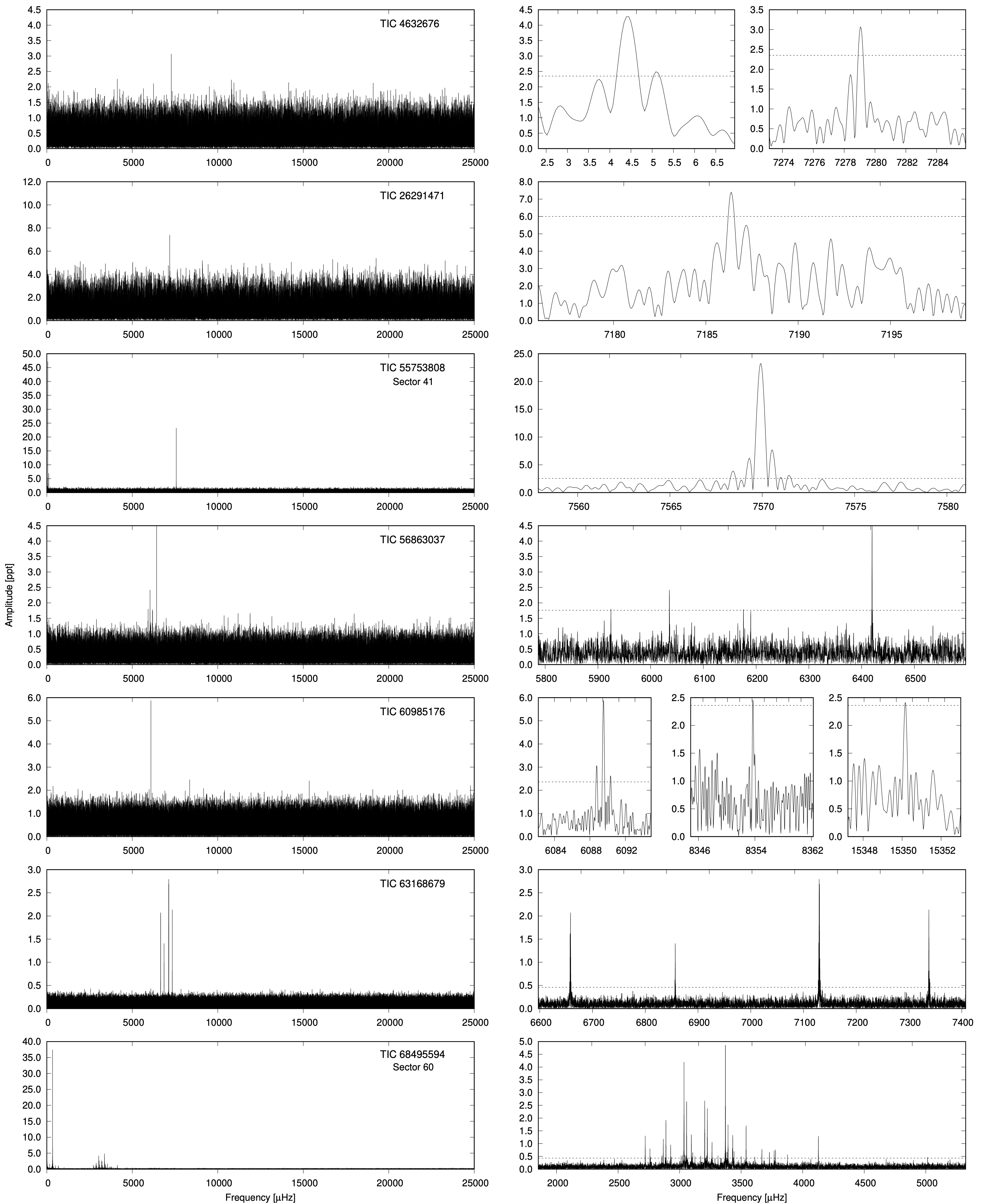}
\caption{Amplitude spectra of targets observed in the USC mode. If sector is not specified, we included all USC data. {\it Left panels:} Frequency range up to the Nyquist frequency. {\it Right panels:} Close-ups of the detected frequencies.}
\label{fig:USC_ft1}
\end{figure*}

For all targets observed in the USC mode, there is also corresponding SC data available. In addition, some targets listed below were observed during specific sectors (1--13) in SC mode as USC was not available. The Nyquist frequency of USC data, 25,000\,\uHz, is well beyond expected pulsation frequencies and so USC provides us with a unique frequency identification. As such we decided not to include any SC data in this paper, even those taken in sectors without available USC data. Table\,\ref{tab:targets_all} provides detailed sector information, including sectors with SC data that were not used in our analyses. We show the amplitude spectra of SC-observed sdBV stars in Figures\,\ref{fig:USC_ft1} through \ref{fig:USC_ft6} and provide the seismic properties in Table\,\ref{tab:USCfreq}.

%\begin{table}{}
%\caption{List of frequencies we detected in the targets observed in the USC mode. The table is available online.}
%\label{tab:USCfreq}
%\end{table}

\onecolumn
\begin{longtable}{cllrr}
\caption{List of frequencies we detected in the targets observed in the USC. If sector is not specified, we included all USC data.}\\
\label{tab:USCfreq}\\
\hline\hline
\multirow{2}{*}{ID} & \multicolumn{1}{c}{Frequency} & \multicolumn{1}{c}{Period} & \multicolumn{1}{c}{Amplitude} & \multirow{2}{*}{S/N}\\
& \multicolumn{1}{c}{[\uHz]} & \multicolumn{1}{c}{[sec]} & \multicolumn{1}{c}{[ppt]} & \\
\hline\hline
\endfirsthead
\caption{continued.}\\
\hline\hline
\multirow{2}{*}{ID} & \multicolumn{1}{c}{Frequency} & \multicolumn{1}{c}{Period} & \multicolumn{1}{c}{Amplitude} & \multirow{2}{*}{S/N}\\
& \multicolumn{1}{c}{[\uHz]} & \multicolumn{1}{c}{[sec]} & \multicolumn{1}{c}{[ppt]} & \\
\hline\hline
\endhead
\hline
\endfoot
\multicolumn{5}{c}{{\bf TIC\,4632676}}\\
$\Omega$ &    4.420(25) &  226244(1283)  &  4.26(44) &   8.1\\
f$_{\rm 1}$   & 7279.068(34) &   137.3802(6)  &  3.08(44) &   5.9\\
\hline
\multicolumn{5}{c}{{\bf TIC\,26291471}}\\
f$_{\rm 1}$   & 7186.374(38) &  139.1522(7)   &  7.4(1)   &   5.6\\
\hline
\multicolumn{5}{c}{{\bf TIC\,55753808\,--\,Sector\,41}}\\
f$_{\rm 1}$ & 7569.914(5) &    132.10189(9) &  23.3(5) &  40.8\\\hline
\multicolumn{5}{c}{{\bf TIC\,56863037}}\\
f$_{\rm 1}$ & 5924.675(41) &    168.7856(12)  &   1.81(33) &   4.6\\
f$_{\rm 2}$ & 6035.540(31) &    165.6853(8)   &   2.42(33) &   6.2\\
f$_{\rm 3}$ & 6175.911(42) &    161.9194(11)  &   1.79(33) &   4.6\\
f$_{\rm 4}$ & 6189.800(43) &    161.5561(11)  &   1.74(33) &   4.5\\
f$_{\rm 5}$ & 6419.795(17) &    155.76822(40) &   4.52(33) &  11.6\\
\hline
\multicolumn{5}{c}{{\bf TIC\,60985176}}\\
f$_{\rm 1}$ &   6089.529(9) &    164.21631(25) &   5.89(45) &  11.2\\
f$_{\rm 2}$ &  8353.720(22) &    119.70715(31) &   2.46(45) &   4.7\\
f$_{\rm 3}$ & 15350.164(22) &     65.14588(9)  &   2.42(45) &   4.6\\
\hline
\multicolumn{5}{c}{{\bf TIC\,63168679}}\\
f$_{\rm 1}$ & 6657.017(10)  &    150.21743(23)  &   0.59(9)  &   5.8\\
f$_{\rm 2}$ & 6657.4928(44) &    150.20670(10)  &   1.44(9)  &  14.0\\
f$_{\rm 3}$ & 6657.976(5)   &    150.19579(12)  &   1.30(10) &  12.7\\
f$_{\rm 4}$ & 6658.2676(44) &    150.18922(10)  &   1.93(10) &  18.8\\
f$_{\rm 5}$ & 6658.629(9)   &    150.18106(20)  &   0.79(10) &   7.8\\
f$_{\rm 6}$ & 6856.7648(41) &    145.84137(9)   &   1.41(9)  &  13.8\\
f$_{\rm 7}$ & 7129.6351(19) &    140.259632(38) &   3.16(9)  &  30.8\\
f$_{\rm 8}$ & 7130.2417(20) &    140.247700(39) &   3.06(9)  &  29.8\\
f$_{\rm 9}$ & 7337.6183(27) &    136.284003(50) &   2.15(9)  &  21.0\\
f$_{\rm 10}$ & 7339.831(10) &    136.24292(18)  &   0.61(9)  &   5.9\\
\hline
\multicolumn{5}{c}{{\bf TIC\,82359147}}\\
f$_{\rm 1}$ & 6891.368(21) &    145.10908(45) &   4.7(7) &   5.4\\
f$_{\rm 2}$ & 7567.595(21) &    132.14237(37) &   4.8(7) &   5.5\\
\hline
\multicolumn{5}{c}{{\bf TIC\,88484868\,--\,Sector\,47}}\\
f$_{\rm 1}$  &  173.607(19)  &  5760.2(6)     &   1.43(12) &  10.1\\
f$_{\rm 2}$  &  180.171(42)  &  5550.3(1.3)   &   0.67(12) &   4.7\\
f$_{\rm 3}$  &  186.959(30)  &  5348.8(9)     &   0.91(12) &   6.4\\
f$_{\rm 4}$  &  216.256(15)  &  4624.15(31)   &   1.90(12) &  13.4\\
f$_{\rm 5}$  &  222.732(18)  &  4489.70(37)   &   1.53(12) &  10.7\\
f$_{\rm 6}$  &  273.006(23)  &  3662.93(30)   &   1.23(12) &   8.7\\
f$_{\rm 7}$  &  291.265(13)  &  3433.30(16)   &   2.09(12) &  14.7\\
f$_{\rm 8}$  &  298.380(11)  &  3351.43(12)   &   2.54(12) &  17.9\\
f$_{\rm 9}$  & 2721.901(5)   &  367.3903(7)   &   5.23(12) &  36.9\\
f$_{\rm 10}$ & 2744.9768(31) &  364.30180(42) &   9.66(14) &  68.1\\
f$_{\rm 11}$ & 2745.647(8)   &  364.2129(11)  &   3.81(15) &  26.8\\
f$_{\rm 12}$ & 2746.415(11)  &  364.1111(14)  &   2.87(14) &  20.2\\
\hline
\multicolumn{5}{c}{{\bf TIC\,88484868\,--\,Sector\,60}}\\
f$_{\rm 1}$  &  123.997(35) &   8064.7(2.3)   &   0.77(11) &   5.9\\
f$_{\rm 2}$  &  173.569(21) &   5761.4(7)     &   1.31(11) &  10.1\\
f$_{\rm 3}$  &  180.323(25) &   5545.6(8)     &   1.08(11) &   8.3\\
f$_{\rm 4}$  &  186.933(34) &   5349.5(10)    &   0.80(11) &   6.2\\
f$_{\rm 5}$  &  216.309(18) &   4623.01(39)   &   1.49(11) &  11.4\\
f$_{\rm 6}$  &  222.719(22) &   4489.96(44)   &   1.24(11) &   9.6\\
f$_{\rm 7}$  &  273.024(32) &   3662.68(42)   &   0.86(11) &   6.6\\
f$_{\rm 8}$  &  291.230(12) &   3433.71(14)   &   2.32(11) &  17.9\\
f$_{\rm 9}$  &  298.360(6)  &   3351.66(7)    &   4.48(11) &  34.5\\
f$_{\rm 10}$ &  317.579(22) &   3148.82(22)   &   1.20(11) &   9.3\\
f$_{\rm 11}$ &  343.986(45) &   2907.10(38)   &   0.61(11) &   4.7\\
f$_{\rm 12}$ & 2721.913(9)  &    367.3887(12) &   3.34(12) &  25.7\\
f$_{\rm 13}$ & 2722.467(25) &    367.3139(34) &   1.22(12) &   9.4\\
f$_{\rm 14}$ & 2745.054(7)  &    364.2916(9)  &  13.27(37) & 102.0\\
f$_{\rm 15}$ & 2745.364(22) &    364.2504(29) &   4.23(36) &  32.5\\
f$_{\rm 16}$ & 2746.307(7)  &    364.1253(9)  &   4.53(15) &  34.9\\
f$_{\rm 17}$ & 3043.387(37) &    328.5813(40) &   0.73(11) &   5.6\\
\hline
\multicolumn{5}{c}{{\bf TIC\,88565376\,--\,Sector\,47}}\\
f$_{\rm 1}$ & 2605.994(20)  &    383.7307(29) &   6.7(6)   &   9.8\\
f$_{\rm 2}$ & 2753.9552(43) &    363.1141(6)  &  30.9(6)   &  45.1\\
\hline
\multicolumn{5}{c}{{\bf TIC\,88565376\,--\,Sector\,60}}\\
f$_{\rm 1}$ & 2605.988(21)  &    383.7317(31) &   5.55(49) &   9.7\\
f$_{\rm 2}$ & 2753.939(6)   &    363.1163(8)  &  19.65(49) &  34.5\\
\hline
\multicolumn{5}{c}{{\bf TIC\,114196505}}\\
f$_{\rm 1}$ &  106.727(40)  &   9369.7(3.5)   &   7.3(1)   &   4.6\\
f$_{\rm 2}$ & 2649.222(9)   &    377.4693(13) &  31.9(1)   &  20.3\\
\hline
\multicolumn{5}{c}{{\bf TIC\,136975077}}\\
f$_{\rm 1}$ & 5045.308(29)  &   198.2040(11)  &   0.88(17) &   5.7\\
f$_{\rm 2}$ & 5045.508(13)  &   198.1961(5)   &   1.94(17) &  12.5\\
f$_{\rm 3}$ & 5084.231(21)  &   196.6866(8)   &   0.71(13) &   4.6\\
f$_{\rm 4}$ & 5093.9751(34) &   196.31034(13) &   4.30(13) &  27.8\\
f$_{\rm 5}$ & 5212.068(12)  &   191.86243(45) &   1.21(13) &   7.8\\
f$_{\rm 6}$ & 5413.021(7)   &   184.73971(23) &   2.23(13) &  14.4\\
f$_{\rm 7}$ & 5481.8018(38) &   182.42177(13) &   3.89(13) &  25.2\\
\hline
\multicolumn{5}{c}{{\bf TIC\,137502282\,--\,Sector\,60}}\\
f$_{\rm 1}$ & 6162.566(17)  &   162.27008(45) &   3.59(25) &  12.2\\
f$_{\rm 2}$ & 7074.937(16)  &   141.34401(32) &   3.91(25) &  13.3\\
\hline
\multicolumn{5}{c}{{\bf TIC\,138618727\,--\,Sector\,41}}\\
$\Omega$  & 33.6846(13) &  29687.1(1.1)  &  23.68(13) & 155.6\\
2$\Omega$ & 67.3692     &  14843.55      &   1.90(13) &  12.5\\
f$_{\rm 1}$ & 2641.979(23)  &   378.5041(34)  &   1.30(13) &   8.6\\
f$_{\rm 2}$ & 2837.514(25)  &   352.4212(31)  &   1.21(13) &   7.9\\
f$_{\rm 3}$ & 2850.814(10)  &   350.7770(12)  &   3.16(13) &  20.8\\
f$_{\rm 4}$ & 2874.318(36)  &   347.9087(43)  &   0.89(13) &   5.8\\
f$_{\rm 5}$ & 2877.1565(29) &   347.56538(35) &  10.88(13) &  71.5\\
f$_{\rm 6}$ & 2878.653(30)  &   347.3847(36)  &   1.04(13) &   6.8\\
f$_{\rm 7}$ & 2906.269(22)  &   344.0838(26)  &   1.41(13) &   9.3\\
\hline
\multicolumn{5}{c}{{\bf TIC\,138618727\,--\,Sector\,48}}\\
$\Omega$  & 33.6826(14) &  29688.9(1.2)  &  23.40(15) & 135.3\\
2$\Omega$ & 67.3652     &  14844.45      &   2.02(15) &  11.7\\
f$_{\rm 1}$ & 2641.965(36)  &   378.506(5)    &   0.92(15) &   5.3\\
f$_{\rm 2}$ & 2837.508(24)  &   352.4220(30)  &   1.38(15) &   8.0\\
f$_{\rm 3}$ & 2874.372(30)  &   347.9021(36)  &   1.14(15) &   6.6\\
f$_{\rm 4}$ & 2877.1608(28) &   347.56486(34) &  12.47(15) &  72.1\\
f$_{\rm 5}$ & 2878.659(23)  &   347.3839(28)  &   1.53(15) &   8.8\\
f$_{\rm 6}$ & 2906.243(22)  &   344.0868(27)  &   1.49(15) &   8.6\\
\hline
\multicolumn{5}{c}{{\bf TIC\,142398823\,--\,Sector\,48\,--\,49}}\\
f$_{\rm 1}$ & 6934.738(17) &    144.20156(36) &   1.32(21) &   5.3\\
f$_{\rm 2}$ & 7771.846(11) &    128.66956(18) &   2.07(21) &   8.4\\
\hline
\multicolumn{5}{c}{{\bf TIC\,159644241}}\\
f$_{\rm 1}$ & 5472.8591(15)  & 182.71985(5)   &   3.4(5)   &   5.4\\
f$_{\rm 2}$ & 5760.1970(5)   & 173.605172(15) &  10.2(5)   &  16.4\\
f$_{\rm 3}$ & 5760.97251(48) & 173.581804(14) &  10.9(5)   &  17.4\\
\hline
\multicolumn{5}{c}{{\bf TIC\,165312944}}\\
f$_{\rm 1}$ & 5806.563(18)  &  172.2189(5)    &   0.48(8)  &   5.3\\
f$_{\rm 2}$ & 6721.456(14)  &  148.77729(31)  &   1.12(10) &  12.5\\
f$_{\rm 3}$ & 6721.592(22)  &  148.77428(50)  &   0.70(10) &   7.8\\
f$_{\rm 4}$ & 6961.3781(16) &  143.649717(32) &   5.54(8)  &  61.5\\
f$_{\rm 5}$ & 7488.717(16)  &  133.53423(28)  &   0.67(9)  &   7.4\\
f$_{\rm 6}$ & 7489.4568(14) &  133.521032(25) &   7.97(10) &  88.6\\
f$_{\rm 7}$ & 7490.125(8)   &  133.50912(14)  &   1.30(9)  &  14.4\\
f$_{\rm 8}$ & 7807.7404(34) &  128.07803(6)   &   2.53(8)  &  28.2\\
f$_{\rm 9}$ & 8169.0575(27) &  122.413142(40) &   3.25(8)  &  36.2\\
\hline
\multicolumn{5}{c}{{\bf TIC\,166054500}}\\
$_{\rm 1}$  & 6680.396(14)  &  149.69173(31)  &   1.00(12) &   6.9\\
\hline
\multicolumn{5}{c}{{\bf TIC\,175402069\,--\,Sector\,46}}\\
f$_{\rm 1}$  & 5298.045(34)  &  188.7489(12)  &   1.16(17) &   6.0\\
f$_{\rm 2}$  & 5392.023(48)  &  185.4592(17)  &   1.43(21) &   7.3\\
f$_{\rm 3}$  & 5392.454(46)  &  185.4443(16)  &   1.50(21) &   7.7\\
f$_{\rm 4}$  & 5417.004(46)  &  184.6039(16)  &   0.87(17) &   4.5\\
f$_{\rm 5}$  & 5435.4698(46) &  183.97674(15) &   8.77(17) &  45.0\\
f$_{\rm 6}$  & 5444.324(21)  &  183.6775(7)   &   1.87(17) &   9.6\\
f$_{\rm 7}$  & 5470.650(37)  &  182.7936(12)  &   1.09(17) &   5.6\\
f$_{\rm 8}$  & 5562.301(27)  &  179.7817(9)   &   1.49(17) &   7.7\\
f$_{\rm 9}$  & 5585.651(25)  &  179.0302(8)   &   1.57(17) &   8.1\\
f$_{\rm 10}$ & 5760.736(16)  &  173.58894(50) &   2.43(17) &  12.4\\
f$_{\rm 11}$ & 7071.184(31)  &  141.4190(6)   &   1.29(17) &   6.6\\
\hline
\multicolumn{5}{c}{{\bf TIC\,175402069\,--\,Sector\,50}}\\
f$_{\rm 1}$  & 5219.087(47)  &  191.6044(17)  &   0.82(16) &   4.5\\
f$_{\rm 2}$  & 5298.072(24)  &  188.7479(9)   &   1.57(16) &   8.6\\
f$_{\rm 3}$  & 5369.360(40)  &  186.2420(14)  &   0.95(16) &   5.2\\
f$_{\rm 4}$  & 5435.4582(28) &  183.97713(9)  &  13.76(16) &  75.2\\
f$_{\rm 5}$  & 5444.320(19)  &  183.6777(6)   &   2.06(16) &  11.2\\
f$_{\rm 6}$  & 5470.662(41)  &  182.7932(14)  &   0.93(16) &   5.1\\
f$_{\rm 7}$  & 5552.316(44)  &  180.1050(14)  &   0.88(16) &   4.8\\
f$_{\rm 8}$  & 5562.355(45)  &  179.7800(15)  &   0.85(16) &   4.7\\
f$_{\rm 9}$  & 5585.631(26)  &  179.0308(8)   &   1.50(16) &   8.2\\
f$_{\rm 10}$ & 5757.213(39)  &  173.6952(12)  &   0.99(16) &   5.4\\
f$_{\rm 11}$ & 5760.722(15)  &  173.58935(47) &   2.48(16) &  13.6\\
f$_{\rm 12}$ & 7071.229(35)  &  141.4181(7)   &   1.10(16) &   6.0\\
\hline
\multicolumn{5}{c}{{\bf TIC\,178081355}}\\
f$_{\rm 1}$ & 6022.331(35)  &  166.0487(10)   &   1.93(30) &   5.5\\
f$_{\rm 2}$ & 6668.937(39)  &  149.9489(9)    &   1.73(30) &   5.0\\
f$_{\rm 3}$ & 6902.357(11)  &  144.87806(24)  &   5.88(30) &  16.8\\
\hline
\multicolumn{5}{c}{{\bf TIC\,191442416}}\\
f$_{\rm 1}$ & 4271.481(28)  &    234.1108(16) &   4.8(7)   &   6.3\\
f$_{\rm 2}$ & 5482.207(20)  &    182.4083(7)  &   6.7(7)   &   8.8\\
\hline
\multicolumn{5}{c}{{\bf TIC\,202354658\,--\,Sector\,48\,--\,51}}\\
f$_{\rm 1}$ & 3074.957(6)   &  325.2078(6)    &   1.45(16) &   7.9\\
f$_{\rm 2}$ & 3077.0481(10) &  324.98680(10)  &   8.96(16) &  48.8\\
f$_{\rm 3}$ & 3079.130(7)   &  324.7670(7)    &   1.31(16) &   7.1\\
f$_{\rm 4}$ & 3088.414(7)   &  323.7908(7)    &   1.26(16) &   6.9\\
f$_{\rm 5}$ & 3090.5031(10) &  323.57191(11)  &   9.36(16) &  51.0\\
f$_{\rm 6}$ & 3091.167(7)   &  323.5024(7)    &   1.43(16) &   7.8\\
f$_{\rm 7}$ & 3091.848(5)   &  323.4312(5)    &   1.81(17) &   9.9\\
f$_{\rm 8}$ & 3092.595(7)   &  323.3530(7)    &   1.30(17) &   7.1\\
\hline
\multicolumn{5}{c}{{\bf TIC\,207440586\,--\,Sector\,56}}\\
f$_{\rm 1}$ & 6945.996(21)  &   143.96784(43) &   2.20(20) &   9.3\\
f$_{\rm 2}$ & 6948.178(29)  &   143.9226(6)   &   1.59(20) &   6.7\\
f$_{\rm 3}$ & 7191.515(31)  &   139.0528(6)   &   1.49(20) &   6.3\\
\hline
\multicolumn{5}{c}{{\bf TIC\,219492314}}\\
f$_{\rm 1}$ & 6936.109(5)   &   144.17305(11) &   5.56(24) &  20.1\\
\hline
\multicolumn{5}{c}{{\bf TIC\,240868270}}\\
f$_{\rm A}$ &    1.7226(34) & 580530(1155)    &  2.212(35) &  59.9\\
f$_{\rm B}$ &    3.4117(15) & 293105(125)     &  5.316(33) & 144.1\\
f$_{\rm C}$ &    6.4641(29) & 154701(69)      &  2.685(32) &  72.8\\
f$_{\rm 1}$ & 6219.191(7)   &   160.79262(18) &  1.096(32) &  29.7\\
f$_{\rm 2}$ & 6220.319(7)   &   160.76347(19) &  1.048(32) &  28.4\\
f$_{\rm 3}$ & 6886.900(21)  &   145.20322(45) &  0.367(33) &   9.9\\
f$_{\rm 4}$ & 6888.509(11)  &   145.16930(24) &  0.692(33) &  18.7\\
f$_{\rm 5}$ & 7128.9861(26) &   140.27240(5)  &  2.911(32) &  78.9\\
\hline
\multicolumn{5}{c}{{\bf TIC\,266013993}}\\
f$_{\rm 1}$ & 5240.491(7)   &   190.82182(27) &   2.08(13) &  14.0\\
f$_{\rm 2}$ & 5284.909(9)   &   189.21803(31) &   1.75(13) &  11.8\\
f$_{\rm 3}$ & 5611.984(14)  &   178.19011(44) &   1.08(13) &   7.3\\
f$_{\rm 4}$ & 7233.366(17)  &   138.24822(32) &   0.90(13) &   6.0\\
\hline
\multicolumn{5}{c}{{\bf TIC\,273255412}}\\
f$_{\rm 1}$  &  200.757(37)  &  4981.1(9)     &   0.59(10) &   5.1\\
f$_{\rm 2}$  & 1767.072(35)  &   565.908(11)  &   0.61(10) &   5.3\\
f$_{\rm 3}$  & 1822.727(7)   &   548.6286(21) &   4.34(12) &  37.7\\
f$_{\rm 4}$  & 1823.391(8)   &   548.4287(24) &   3.81(12) &  33.2\\
f$_{\rm 5}$  & 1883.332(7)   &   530.9737(19) &   3.65(10) &  31.8\\
f$_{\rm 6}$  & 1884.891(37)  &   530.535(10)  &   0.67(10) &   5.8\\
f$_{\rm 7}$  & 1905.196(28)  &   524.880(8)   &   0.78(10) &   6.8\\
f$_{\rm 8}$  & 1934.090(7)   &   517.0389(17) &   3.60(12) &  31.3\\
f$_{\rm 9}$  & 1934.7020(35) &   516.8755(9)  &   6.71(12) &  58.4\\
f$_{\rm 10}$ & 1972.958(24)  &   506.853(6)   &   0.90(10) &   7.8\\
f$_{\rm 11}$ & 2059.224(7)   &   485.6199(16) &   3.33(12) &  29.0\\
f$_{\rm 12}$ & 2060.167(21)  &   485.3975(50) &   1.08(12) &   9.4\\
f$_{\rm 13}$ & 2141.805(8)   &   466.8960(17) &   2.73(10) &  23.8\\
f$_{\rm 14}$ & 2214.0914(32) &   451.6525(7)  &   7.73(11) &  67.2\\
f$_{\rm 15}$ & 2215.170(18)  &   451.4326(36) &   1.40(11) &  12.1\\
f$_{\rm 16}$ & 2329.046(13)  &   429.3604(24) &   1.65(10) &  14.4\\
f$_{\rm 17}$ & 2392.977(19)  &   417.8896(33) &   1.14(10) &   9.9\\
f$_{\rm 18}$ & 2473.978(30)  &   404.2073(49) &   0.72(10) &   6.3\\
f$_{\rm 19}$ & 2498.3133(39) &   400.2700(6)  &   5.57(10) &  48.5\\
f$_{\rm 20}$ & 2520.811(31)  &   396.6978(48) &   0.92(12) &   8.0\\
f$_{\rm 21}$ & 2521.894(37)  &   396.527(6)   &   0.82(12) &   7.1\\
f$_{\rm 22}$ & 2523.414(41)  &   396.289(6)   &   0.66(12) &   5.7\\
f$_{\rm 23}$ & 2524.833(43)  &   396.066(7)   &   0.54(11) &   4.7\\
f$_{\rm 24}$ & 2593.443(36)  &   385.588(5)   &   0.60(10) &   5.2\\
f$_{\rm 25}$ & 2658.533(39)  &   376.147(5)   &   1.01(12) &   8.7\\
f$_{\rm 26}$ & 2659.046(25)  &   376.0747(35) &   1.56(12) &  13.5\\
\hline
\multicolumn{5}{c}{{\bf TIC\,284692897\,--\,Sector\,40\,--\,41}}\\
$\Omega$ &  121.7120(7) &   8216.115(48) &   1.06(19) &   4.7\\
2$\Omega$ & 243.4240    &   4108.0575    &  15.27(19) &  67.5\\
f$_{\rm 1}$  & 3665.134(15)  &   272.8413(11) &   1.40(19) &   6.2\\
f$_{\rm 2}$  & 3733.129(10)  &    267.8718(7) &   2.14(19) &   9.5\\
f$_{\rm 3}$  & 3792.730(14)  &   263.6623(10) &   1.52(19) &   6.7\\
f$_{\rm 4}$  & 3823.710(15)  &   261.5261(10) &   1.48(19) &   6.6\\
f$_{\rm 5}$  & 3885.889(18)  &   257.3414(12) &   1.21(19) &   5.3\\
f$_{\rm 6}$  & 3908.570(5)   &  255.84805(34) &   4.21(19) &  18.6\\
f$_{\rm 7}$  & 3924.770(18)  &   254.7920(12) &   1.17(19) &   5.2\\
f$_{\rm 8}$  & 3933.711(18)  &   254.2129(11) &   1.22(19) &   5.4\\
f$_{\rm 9}$  & 3992.693(21)  &   250.4575(13) &   1.01(19) &   4.5\\
f$_{\rm 10}$ & 4053.126(21)  &   246.7231(13) &   1.01(19) &   4.5\\
f$_{\rm 11}$ & 4129.329(13)  &    242.1701(8) &   1.63(19) &   7.2\\
f$_{\rm 12}$ & 4168.217(17)  &   239.9108(10) &   1.24(19) &   5.5\\
f$_{\rm 13}$ & 4177.108(15)  &    239.4001(8) &   1.48(19) &   6.6\\
f$_{\rm 14}$ & 4179.068(21)  &   239.2878(12) &   1.05(19) &   4.7\\
\hline
\multicolumn{5}{c}{{\bf TIC\,309807601}}\\
f$_{\rm 1}$  & 6896.776(17)  &  144.99528(36) &   1.87(28) &   5.8\\
f$_{\rm 2}$  & 7344.805(11)  &  136.15066(21) &   2.87(28) &   8.8\\
\hline
\multicolumn{5}{c}{{\bf TIC\,310937915}}\\
f$_{\rm 1}$  & 6963.373(33)  &    143.6086(7) &  11.0(2)   &   6.2\\
\hline
\multicolumn{5}{c}{{\bf TIC\,355754830}}\\
f$_{\rm 1}$  &  275.756(20)  &  3626.39(26)   &   0.80(7) &  10.3\\
f$_{\rm 2}$  &  403.764(25)  &  2476.69(15)   &   0.64(7) &   8.3\\
f$_{\rm 3}$  &  453.478(37)  &  2205.18(18)   &   0.43(7) &   5.5\\
f$_{\rm 4}$  &  565.832(10)  &  1767.310(32)  &   1.53(7) &  19.8\\
f$_{\rm 5}$  & 3832.835(20)  &  260.9035(13)  &   0.81(7) &  10.5\\
f$_{\rm 6}$  & 3841.620(9)   &  260.3069(6)   &   2.24(8) &  28.8\\
f$_{\rm 7}$  & 3842.4635(27) &  260.24971(18) &   7.24(9) &  93.4\\
f$_{\rm 8}$  & 3843.2473(26) &  260.19663(17) &   6.72(9) &  86.7\\
f$_{\rm 9}$  & 3845.345(19)  &  260.0547(13)  &   0.91(7) &  11.7\\
f$_{\rm 10}$ & 3846.846(8)   &  259.9533(5)   &   2.59(9) &  33.4\\
f$_{\rm 11}$ & 3847.491(9)   &  259.9096(6)   &   2.50(9) &  32.3\\
f$_{\rm 12}$ & 3848.369(20)  &  259.8503(13)  &   0.94(8) &  12.1\\
f$_{\rm 13}$ & 3874.494(37)  &  258.0982(25)  &   0.42(7) &   5.5\\
f$_{\rm 14}$ & 4028.902(19)  &  248.2066(12)  &   0.84(7) &  10.8\\
\hline
\multicolumn{5}{c}{{\bf TIC\,357232133\,--\,Sector\,47}}\\
f$_{\rm 1}$  & 6744.690(31)  &    148.2648(7) &  1.98(27) &   6.3\\
f$_{\rm 2}$  & 7167.391(8)   &  139.52078(16) &  7.21(27) &  22.9\\
f$_{\rm 3}$  & 8065.521(16)  &  123.98455(24) &  4.00(27) &  12.7\\
f$_{\rm 4}$  & 8066.831(30)  &  123.96441(45) &  2.10(27) &   6.7\\
\hline
\multicolumn{5}{c}{{\bf TIC\,364966239\,--\,Sectors\,54\,--\,57,60}}\\
f$_{\rm 1}$  & 6568.4587(38) &  152.24272(9)  &  0.91(10) &   7.8\\
f$_{\rm 2}$  & 6569.0255(17) & 152.229581(40) &  2.08(10) &  17.7\\
f$_{\rm 3}$  & 6569.608(6)   &  152.21609(15) &  0.54(10) &   4.6\\
f$_{\rm 4}$  & 6787.6974(24) &  147.32537(5)  &  1.40(10) &  11.9\\
\hline
\multicolumn{5}{c}{{\bf TIC\,397595169}}\\
f$_{\rm 1}$  & 2824.115(33)  &   354.0932(41) &  2.08(30) &   5.8\\
f$_{\rm 2}$  & 2860.889(33)  &   349.5417(40) &  2.09(30) &   5.9\\
\hline
\multicolumn{5}{c}{{\bf TIC\,424720852\,--\,Sectors\,40\,--\,41}}\\
f$_{\rm 1}$  &  105.593(13)  &   9470.3(1.2)  &   0.62(7) &   7.4\\
f$_{\rm 2}$  & 2469.440(17)  &   404.9501(28) &   0.51(7) &   6.1\\
f$_{\rm 3}$  & 2469.752(15)  &   404.8990(25) &   0.58(7) &   6.9\\
f$_{\rm 4}$  & 2470.079(8)   &   404.8453(13) &   1.06(7) &  12.8\\
f$_{\rm 5}$  & 2662.554(18)  &   375.5792(25) &   0.45(7) &   5.4\\
f$_{\rm 6}$  & 2680.773(5)   &   373.0268(7)  &  2.87(10) &  34.4\\
f$_{\rm 7}$  & 2681.0261(41) &   372.9915(6)  &   5.78(8) &  69.3\\
f$_{\rm 8}$  & 2681.289(15)  &   372.9550(20) &   1.11(9) &  13.3\\
f$_{\rm 9}$  & 2708.836(18)  &   369.1622(25) &   0.44(7) &   5.3\\
\hline
\multicolumn{5}{c}{{\bf TIC\,424720852\,--\,Sectors\,50,54\,--\,57,60}}\\
f$_{\rm 1}$  &  105.6188(34)  &   9468.01(30)  & 0.328(46) &   6.1\\
f$_{\rm 2}$  &  175.7895(27)  &   5688.62(9)   & 0.408(46) &   7.6\\
f$_{\rm 3}$  &  271.0415(46)  &   3689.47(6)   & 0.239(46) &   4.5\\
f$_{\rm 4}$  &  289.9835(36)  &  3448.472(43)  & 0.482(50) &   9.0\\
f$_{\rm 5}$  &  290.066(6)    &   3447.49(7)   & 0.308(50) &   5.8\\
f$_{\rm 6}$  &  309.8852(31)  &  3227.001(33)  & 0.353(46) &   6.6\\
f$_{\rm 7}$  & 2469.5573(31)  &  404.9309(5)   & 0.398(50) &   7.4\\
f$_{\rm 8}$  & 2469.6820(17)  & 404.91043(28)  & 0.727(50) &  13.6\\
f$_{\rm 9}$  & 2470.0434(11)  & 404.85119(18)  & 0.997(46) &  18.6\\
f$_{\rm 10}$ & 2662.5412(18)  & 375.58104(25)  & 0.620(46) &  11.6\\
f$_{\rm 11}$ & 2680.76199(49) &  373.02827(7)  & 2.531(48) &  47.3\\
f$_{\rm 12}$ & 2681.02654(22) & 372.991458(30) & 5.852(47) & 109.4\\
f$_{\rm 13}$ & 2681.2944(10)  &  372.95420(14) & 1.209(48) &  22.6\\
f$_{\rm 14}$ & 2708.8358(30)  &  369.16228(41) & 0.367(46) &   6.9\\
\hline
\multicolumn{5}{c}{{\bf TIC\,441725813\,--\,Sectors\,55\,--\,60}}\\
f$_{\rm 1}$  & 2755.5369(38) &   362.9057(5)   &  0.098(9) &  11.1\\
f$_{\rm 2}$  & 2768.4702(24) &   361.21032(31) &  0.162(9) &  18.2\\
f$_{\rm 3}$  & 2769.1459(32) &   361.12218(41) &  0.122(9) &  13.7\\
f$_{\rm 4}$  & 2769.817(6)   &   361.0346(8)   &  0.066(9) &   7.4\\
f$_{\rm 5}$  & 2797.018(7)   &   357.5236(9)   &  0.054(9) &   6.1\\
f$_{\rm 6}$  & 2798.413(9)   &   357.3454(12)  &  0.041(9) &   4.6\\
f$_{\rm 7}$  & 3660.781(8)   &   273.1658(6)   &  0.049(9) &   5.5\\
f$_{\rm 8}$  & 3661.141(6)   &   273.13886(46) &  0.061(9) &   6.9\\
f$_{\rm 9}$  & 3674.845(8)   &   272.1203(6)   &  0.046(9) &   5.1\\
f$_{\rm 10}$ & 3678.276(7)   &   271.86649(49) & 0.059(10) &   6.7\\
f$_{\rm 11}$ & 3678.394(6)   &   271.85779(42) & 0.068(10) &   7.7\\
f$_{\rm 12}$ & 3678.593(6)   &   271.84307(42) & 0.067(10) &   7.6\\
f$_{\rm 13}$ & 3680.469(7)   &   271.7045(5)   & 0.056(10) &   6.3\\
f$_{\rm 14}$ & 3680.617(7)   &   271.69360(50) & 0.058(10) &   6.5\\
\hline
\multicolumn{5}{c}{{\bf TIC\,471015194}}\\
f$_{\rm 1}$  & 5629.2847(16) &  177.64246(5)   &  1.22(20) &   5.2\\
f$_{\rm 2}$  & 5641.0876(10) &  177.270779(33) &  1.90(20) &   8.1\\
f$_{\rm 3}$  & 5788.6736(6)  &  172.751148(18) &  3.31(20) &  14.0\\
f$_{\rm 4}$  & 6025.1463(12) &  165.971073(34) &  1.62(20) &   6.9\\
f$_{\rm 5}$  & 6032.7508(7)  &  165.761862(18) &  3.02(20) &  12.8\\
\hline\hline
\end{longtable}
\twocolumn

\paragraph{TIC\,4632676}
(Ton\,74) is a new sdOB pulsator. The star was classified as an sdB star in the Palomar-Green (PG) survey \citep{green86}. Then, the star was found to be a binary system consisting of sdB and dM stars by \citet{luo16}. Recently it was reclassified as an sdOB star by \citet{lei19a}. \tess\ observed the star during Sector\,49. We detected two frequencies, one in the very low frequency region, which can be interpreted as a binary frequency (orbital period of 2.619(15)\,days), and one in the high frequency region that we associate with a p-mode. We show the amplitude spectrum in Fig.\,\ref{fig:USC_ft1}, and list the prewhitened frequencies in Table\,\ref{tab:USCfreq}.

\paragraph{TIC\,26291471}
(HE\,1450-0957) is a known sdOB pulsator. \citet{lisker05} derived its atmospheric parameters indicative of an sdOB type and \citet{ostensen10a} detected two high frequency signals. \tess\ observed the star during Sector\,51. We detected only one frequency in the high frequency region that we associated with a p-mode. It is close but does not overlap with frequency $f_1$ reported by \citet{ostensen10a}. We show the amplitude spectrum in Fig.\,\ref{fig:USC_ft1}, and list the prewhitened frequency in Table\,\ref{tab:USCfreq}.

\paragraph{TIC\,55753808}
(PG\,1203+574) is a new sdOB pulsator. The star was originally classified as an sdB by \citet{green86}. \citet{lei19b} derived spectroscopic parameters and classified the star as an sdOB. \tess\ observed the star during Sectors\,41 and 48. These two data sets are many sectors apart and we decided not to merge both sets, but to analyze them separately. We detect the same single p-mode frequency in each data set. The amplitude spectra also show low frequencies, which are characteristic of binary frequencies. Our contamination analysis revealed that the binary signal comes from neighboring TIC\,55753802, which is a known binary star and we provide no further analysis of this signal. We show the amplitude spectrum for Sector\,41 data in Fig.\,\ref{fig:USC_ft1}, and list the prewhitened frequency in Table\,\ref{tab:USCfreq}.

\paragraph{TIC\,56863037}
(DT\,Lyn) is a known sdB pulsator. The star was originally classified as an sdB-O by \citet{green86}. \citet{koen99} obtained photometry and a spectrum, which allowed them to sharpen the classification to sdB, and detect three p-modes. \citet{randall07} independently classified the star as an sdB and reported seven p-modes. They performed an asteroseismic analysis and delivered structural parameters of the star, e.g. $\log$M$_{\rm env}$/M$_*$\,=\,-4.69$\pm$0.07, M$_*$\,=\,0.39$\pm$0.01\,M$_\odot$, R/R$_\odot$\,=\,0.133$\pm$0.001. \tess\ observed the star during Sector\,47. We detected five frequencies out of which four were previously reported by \citet{randall07}. Frequency f$_3$ was not detected before, however we measured it at S/N\,<\,5 so it is only a tentative detection. We show the amplitude spectrum calculated from all data in Fig.\,\ref{fig:USC_ft1}, and we list the prewhitened frequencies in Table\,\ref{tab:USCfreq}.

\paragraph{TIC\,60985176}
(BI\,Ari) is a known sdOB pulsator. It was originally classified as an sdB-O by \citet{green86}. \citet{koen04} constrained the classification to sdB with a G0V\,--\,G2V companion and reported a discovery of one p-mode with a period near 164\,s. \citet{reed06} confirmed this period and detected another five. Our fit to a spectrum taken with the NOT in 2022 gives \teff\,=\,35\,488(272)\,K, \loggcms\,=\,5.49(5) and \logy\,=\,-1.74(9), and we classified the star as an sdOB. We did the fit assuming the object is a single star. \tess\ observed the star during Sectors\,42 and 43 and we detected three frequencies, the same one as reported by \citet{koen04} and later by \citet{reed06}, which continues to be the highest-amplitude pulsation in this star. The other two frequencies should be considered tentative. Our f$_2$ (8353.72\,\uHz) is 8.37\,\uHz away from \citet{reed06}'s f$_2$ (8362.09\,\uHz) and our very high-frequency f$_3$ (15350.16\,\uHz) is near-to twice \citet{reed06}'s f$_3$ if the latter is off by a daily alias, which could be expected from single-site ground-based data. While \citet{reed06} also used 20\,s integrations, they did not investigate past 10\,000\uHz. We show the amplitude spectrum in Fig.\,\ref{fig:USC_ft1}, and list the prewhitened frequencies in Table\,\ref{tab:USCfreq}.

\paragraph{TIC\,63168679}
(\gaia\,DR3 675213084211549696) is a new sdOB pulsator. The star was classified as an sdOB by \citet{lei20}. \tess\ observed the star during Sectors\,44 and 46 from which we detected 10 frequencies. We show the amplitude spectrum in Fig.\,\ref{fig:USC_ft1}, and list the prewhitened frequencies in Table\,\ref{tab:USCfreq}. We detect possible multiplets around 6660 and 7340\,\uHz. The frequency spacing between f$_1$ and f$_3$ is comparable to the one between f$_7$ and f$_8$. However, f$_2$ is not located equidistant from f$_1$ and f$_3$. If this is a triplet it is very asymmetric. Higher frequency resolution is required to make a definite statement whether these are multiplets but if confirmed, the rotation period would be around 10.5\,days.

\paragraph{TIC\,68495594}
(HD\,265435) is a known sdOB pulsator. This star has been extensively studied prior to our analysis by \citet{pelisoli21} and \citet{jayaraman22}. Our fit to a spectrum taken with the NOT in 2018 gives \teff\,=\,33\,253(137)\,K, \loggcms\,=\,5.45(2) and \logy\,=\,-1.40(5); the exposure time of 150 seconds is short compared to the 1.65\,hr orbital period, implying minimal effects of orbital line broadening. \tess\ observed the star during Sectors\,44, 45 and 60. The latter authors analyzed only the first two sectors mentioned, while data collected during Sector\,60 were not analyzed by them. We show an amplitude spectrum for that sector in Fig.\,\ref{fig:USC_ft1}. It does not vary from the previous sectors significantly and we decided not to present any solution in Table\,\ref{tab:USCfreq}. Since the star shows tidally tilted pulsations we expect some modes can come and go with time, however the overall picture will not change. Sector\,60's data set can surely be a useful check on the model presented by \citet{jayaraman22}, however that is beyond the scope of this paper.

\paragraph{TIC\,82359147}
(PG\,1657+416) is a known sdB pulsator. It was classified as an sdB star by \citet{green86}. \citet{oreiro07} confirmed the spectral classification, detected a signature of a G5 companion, and discovered high frequency signals. \tess\ observed the star during Sectors\,51 and 52. We detected two frequencies, both previously reported by \citet{oreiro07}. We show the amplitude spectrum in Fig.\,\ref{fig:USC_ft2}, and list the prewhitened frequencies in Table\,\ref{tab:USCfreq}.

\paragraph{TIC\,88484868}
(FBS\,0658+627) is a new sdB pulsator. \citet{nemeth12} classified the star as an sdB. \tess\ observed it during Sectors\,47 and 60. Since these two sectors are too far apart in time, we decided not to merge data for a Fourier analysis. We detected 12 and 17 frequencies, in both the g- and p-mode regions, in Sector 47 and 60, respectively. Although the p-mode frequencies are amplitude dominant, we detected more frequencies in the g-mode region. While the amplitudes change between sectors, most of the frequencies are detected in both sectors. f$_1$--f$_9$ from Sector 47 are recovered in Sector 60, but surprisingly, the frequencies of the highest-amplitude three near 2745\uHz\ are different with the dominant amplitude nearly doubling in Sector 60. These three frequencies may also be a rotationally split triplet, but the frequency splitting is not symmetric and it changes between sectors, making this interpretation less likely. We show the amplitude spectrum for Sector\,60 data in Fig.\,\ref{fig:USC_ft2}, and list the prewhitened frequencies in Table\,\ref{tab:USCfreq}.

\paragraph{TIC\,88565376}
(DW\,Lyn) is a known sdB pulsator. It was classified as an sdB, based on spectroscopic parameters derivation, by \citet{dreizler02}. The authors detected periodic flux variations and interpreted them by stellar pulsations. The pulsation periods were estimated around 363 and 382\,s. Later, \citet{schuh06} collected additional data and confirmed the two periods reported by \citet{dreizler02} and discovered a new one in the g-mode region. The authors explained this newly detected signal as stellar pulsation, making it the first known hybrid sdBV star. \tess\ observed the star during Sectors\,47 and 60. We only detect the two p-mode frequencies reported in previous papers. We show the amplitude spectrum calculated from Sector\,47 data in Fig.\,\ref{fig:USC_ft2}, and we list the prewhitened frequencies in Table\,\ref{tab:USCfreq}. We did not detect the low frequency signal, although if its amplitude remained fairly constant over time, it would be clearly detected in our analysis. In fact, the p-mode amplitudes of Sector\,60 are nearly the same as reported by \cite{schuh06} for 2004 data. Therefore, we might expect the g-mode amplitude to be roughly preserved, though it appears it has not.

\paragraph{TIC\,90960668}
(V585\,Peg, Balloon\,090100001) is a known sdB pulsator. This star has been extensively studied prior to our analysis by \citet{oreiro04,baran09,telting06,vangrootel08a} and others. \tess\ observed the star during Sector\,56. These data are published by \citet{reed23} and hence we only show the amplitude spectrum in Fig.\,\ref{fig:USC_ft2} and refer readers to the latter paper for details on results of \tess\ data.

\paragraph{TIC\,114196505}
(\gaia\,DR3 4187241720861658624) is a new sdB pulsator. Our fit to a spectrum taken with the NOT in 2022 gives \teff\,=\,27\,594(442)\,K, \loggcms\,=\,5.30(6) and \logy\,=\,-2.99(19), and we classified the star as an sdB. \tess\ observed the star during Sector\,54. We detect two frequencies, which we list in Table\,\ref{tab:USCfreq} and show the amplitude spectrum in Fig.\,\ref{fig:USC_ft2}. The frequency in the g-mode region has low S/N and should be confirmed.

\paragraph{TIC\,136975077}
(V2203\,Cyg) is a known sdB pulsator. It was classified as an sdB by \citet{downes86}. \citet{billeres98} derived an independent estimate of spectroscopic parameters and confirmed the sdB type and detected five pulsation frequencies in photometry. \cite{koen98a} re-observed the star and increased the number of detected frequencies to seven. \citet{jeffery00} used spectroscopic data to measure radial velocities of pulsations and detected two frequencies that were previously reported from both photometric analyses. \citet{zhou06} analyzed multi-site photometric data and detected six out of seven frequencies reported by \citet{koen98a}. The missing one was the lowest amplitude frequency in \citet{koen98a}'s analysis. \tess\ observed the star during Sectors\,55 and 56. We detected seven frequencies. We show the amplitude spectrum in Fig.\,\ref{fig:USC_ft2}, and we list the prewhitened frequencies in Table\,\ref{tab:USCfreq}. Comparing our detection with the one presented by \citet{koen98a}, there is a difference in detecting close frequencies. \citet{koen98a} found two close frequencies at 5413\,\uHz, while we found only one in that region. On the other hand we found an additional frequency close to the highest amplitude one, f$_2$ and f$_1$, respectively. We stress, however, that the frequency spacing between these two peaks is comparable to the resolution, so f$_2$ may just be a residual signal caused by unstable frequency/amplitude of f$_1$. Our detection of seven frequencies with an amplitude order very similar to that reported more than 20 years ago, means that the pulsation mechanism remained stable in this star.

\paragraph{TIC\,137502282}
(FBS\,0844+792) is a new sdOB pulsator. Our fit to a spectrum taken with the INT in 2019 gives \teff\,=\,34\,259(267)\,K, \loggcms\,=\,5.79(5) and \logy\,=\,-1.31(7), indicating an sdOB classification. \tess\ observed the star during Sectors\,40, 47, 53 and 60. We detect the same two frequencies of comparable amplitudes in all sectors. While the frequencies remain significant in all four sectors, the amplitude varies significantly i.e. 2.1, 3.01, 1.47 and 3.59\,ppt, 3.76, 3.32, 3.66 and 3.91\,ppt, for Sectors, 40, 47, 53, 60 and the lower and higher frequencies, respectively. Since merging all data together leads to a complex window function without detecting additional pulsations, we only present results from Sector\,60 data analysis. We show the amplitude spectrum in Fig.\,\ref{fig:USC_ft2}, and list the prewhitened frequencies in Table\,\ref{tab:USCfreq}.

\paragraph{TIC\,138618727}
(KL\,UMa, Feige\,48) is a known and well-studied sdB pulsator. \citet{graham70} was the first who classified it as an sdB star. Spectroscopic estimates were reported by \citet{koen98b} who confirmed an sdB type, and discovered five frequencies associated with short-period pulsations modes. \citet{reed04} collected multi-site and multi-year photometry and reported five frequencies although some were daily aliases away from those reported by \citet{koen98b}. \citet{reed04} presumed three of the frequencies formed an asymmetric triplet and predicted a rotation period which was subsequently found to match a binary period with a white dwarf companion discovered by \citet{otoole04}. \citet{reed04} also examined pulsation stability, deriving an upper limit evolutionary scale of $\dot{P}/P=4.9\pm 5.3\times 17^{-16}s^{-1}$. \citet{charpinet05a} reported four frequencies and their results of period fitting, which provided physical parameters of the star, e.g. $\log$M$_{\rm env}$/M$_*$\,=\,-2.97$\pm$0.09, M$_*$\,=\,0.460$\pm$0.008\,M$_\odot$, R/R$_\odot$\,=\,0.2147$\pm$0.0034. \citet{vangrootel08} tested a spin-orbit synchronism and concluded that KL\,UMa rotates as a solid body in a tidally locked system. Subsequently, \citet{reed12c} obtained time-resolved spectroscopy and additional photometry, which slightly shortened the binary period to 0.342(8)\,d, making the rotation (0.418\,d) slightly subsynchronous. \tess\ observed the star in Sectors\,41 and 48. The amplitude spectra in each sector do not contain the exactly the same signal and we decided not to merge data from both sectors. We show the amplitude spectrum for Sector\,41 data in Fig.\,\ref{fig:USC_ft3}, and list the prewhitened frequencies in Table\,\ref{tab:USCfreq}. We detect two low-frequency signals, which we interpret as a binary frequency and its harmonic. With \tess's precision, the orbital solution  derived from combined two sectors becomes 0.3436074(8)\,d in agreement with the \citet{reed12c} value. These signals are the first detection of the binary signal in photometric data. In addition, we detected seven and six high-frequency signals, in Sector\,41 and 48, respectively. Frequency f$_3$ detected in the former data set is not significant in the latter one. Frequency 2877.16\,\uHz\,seems to be accompanied by two low amplitude sidelobe frequencies, one on each side. The frequency spacing is not equal on both sides so it does not call for a rotationally split modes, although the lower-frequency signal is spaced twice as much as the higher-frequency signal. It is still possible the three frequencies are part of a quintuplet with a frequency splitting of 1.5\,\uHz. This would lead to a rotation period of 7.7\,d, which is not consistent with previous conclusions.

\paragraph{TIC\,142398823}
(PG\,1315+645) is a new sdOB pulsator. It was classified as an sdOB+MS by \citet{geier17}. \tess\ observed the star during Sectors\,41, 48 and 49. The window function of the combined data is very complex, which prohibits a unique frequency selection for prewhitening. Therefore, we decided to use only combined Sectors\,48 and 49 and we detected two frequencies. We show the amplitude spectrum in Fig.\,\ref{fig:USC_ft3}, and list the prewhitened frequencies in Table\,\ref{tab:USCfreq}.

\paragraph{TIC\,159644241}
(KIC\,10139564) is a known sdB pulsator. It was classified as an sdB star by \citet{ostensen10b} and extensively observed during the \emph{Kepler} mission. Those data were extensively studied \citep{kawaler10,baran12,baran13,zong16}, making it probably the best-solved p-mode sdBV star and we refer readers to those papers which contain substantial analyses. \tess\ observed the star during Sectors \,40, 41, 54 and 55. We detected three frequencies, two in the previously identified triplet, i.e f$_{33}$ and f$_{35}$ listed by \citet{baran12} and one identified as a singlet, i.e. f$_{18}$. We show the amplitude spectrum in Fig.\,\ref{fig:USC_ft3}, and list the prewhitened frequencies in Table\,\ref{tab:USCfreq}.

\paragraph{TIC\,165312944}
(KY\,UMa) is a known sdB pulsator. Based on spectroscopic parameters \citet{koen99} classified it as an sdB star. The authors also reported four pulsation periods. \citet{charpinet05b} detected nine frequencies, including four reported by \citet{koen99} and derived structural parameters of the star using asteroseismology, e.g. $\log$M$_{\rm env}$/M$_*$\,=\,-4.254$\pm$0.147, M$_*$\,=\,0.457$\pm$0.012\,M$_\odot$, R/R$_\odot$\,=\,0.1397$\pm$0.0028. \citet{reed09} obtained time-resolved spectroscopy, detected four previously known frequencies and low-amplitude occasional transients. They inferred temperature and gravity changes for the four main frequencies. \tess\ observed the star during Sectors\,48 and 49. We detected nine frequencies including four main ones, originally reported by \citet{koen99}. The amplitudes vary over time and during \tess\ monitoring a signal at 7490\,\uHz\ became the highest amplitude one. In fact, this signal seems to be split into three components separated slightly asymmetrically by 0.74 and 0.69\,\uHz. If the signal was a rotationally split mode the average rotation period would equal 16.3\,d. We do not see other peaks indicative of multiplets. We show the amplitude spectrum for Sector\,48 and 49 data in Fig.\,\ref{fig:USC_ft3}, and list the prewhitened frequencies in Table\,\ref{tab:USCfreq}.

\paragraph{TIC\,166054500}
(PG\,1409+605) is a new sdB pulsator. \citet{green86} classified it as an sdB-O. \citet{geier17} derived an sdOB classification and a hint of a main sequence companion. Based on a spectrum taken with the NOT in 2022 we derived an sdB+F classification but the quality of the spectrum is not sufficient to derive a reliable spectroscopic fit. \tess\ observed the star during Sectors\,48 and 49. We detected only one frequency, which we list in Table\,\ref{tab:SCfreq}, while we showed the amplitude spectrum in Fig.\,\ref{fig:USC_ft3}.

\paragraph{TIC\,175402069}
(NY\,Vir, PG\,1336$-$018) is a known sdB pulsator. The star was found to be an eclipsing binary of an HW\,Vir type with a pulsating sdB primary by \citet{kilkenny98}. The authors reported 0.1\,d orbital period and two short period flux variations associated with stellar pulsations. The system became a subject of very intense investigation ever since. The analysis commonly undertaken was related to eclipse timings to search for additional companions to the system, and what is more essential with respect to this paper, the asteroseismic investigation to derive structural properties of the primary. \citet{kilkenny03} reported a list of 28 pulsation periods detected in photometric data collected during a multi-site campaign and \citet{reed01} found the first evidence for tidally-tipped pulsations for an sdBV star, aligning and precessing with the companion. \citet{charpinet08} used the \citet{kilkenny03} list for an asteroseismic analysis and derived parameters of the primary sdB star, e.g. $\log$M$_{\rm env}$/M$_*$\,=\,-4.54$\pm$0.07, M$_*$\,=\,0.459$\pm$0.005\,M$_\odot$, R/R$_\odot$\,=\,0.151$\pm$0.001. \tess\ observed the star during Sectors\,46 and 50. The data were first cleaned of the binary flux variation. The amplitude spectra calculated from both sectors are different and we decided not to merge both data sets together. We detected 11 and 12 frequencies in Sector\,46 and 50, respectively. We skipped signals at an orbital frequency from the highest amplitude frequency. We considered these signals to be artifacts caused by a pulsation phase change during eclipses. Four frequencies were not reported before, i.e. f$_1$, f$_4$, f$_8$, and f$_2$, f$_7$ in Sector\,46 and 50, respectively. Frequencies f$_4$ and f$_7$ should be confirmed with more precise data. We show the amplitude spectrum for Sector\,46 data in Fig.\,\ref{fig:USC_ft3}, and list the prewhitened frequencies in Table\,\ref{tab:USCfreq}. 

\paragraph{TIC\,178081355}
(FBS\,0315+417) is a new sdOB pulsator. It was classified as an sdOB by \citet{geier17}. \tess\ observed the star during Sector\,58. We detected three frequencies. We show the amplitude spectrum in Fig.\,\ref{fig:USC_ft3}, and  list the prewhitened frequencies in Table\,\ref{tab:USCfreq}.

\paragraph{TIC\,191442416}
(V429\,And) is a known sdB pulsator. It was classified as an sdB star by \citet{ostensen01a}. The authors reported a detection of four pulsation periods. \citet{reed07a} collected new photometry and detected 14 frequencies. \tess\ observed the star during Sector\,57. We detected two frequencies, both already reported by \citet{reed07a}. We show the amplitude spectrum in Fig.\,\ref{fig:USC_ft4}, and list the prewhitened frequencies in Table\,\ref{tab:USCfreq}.

\paragraph{TIC\,202354658}
(PG\,1544+601) is a new sdB pulsator. It was identified as an sdB star by \citet{green86}. \tess\ observed the star during Sectors\,48\,--\,51 and 58. Since Sector\,58 data are several months apart from the bulk of continuous data, and this stand alone data set does not allow for detecting additional frequencies, we decided to analyze only data collected during Sectors\,48\,--\,51. We detected eight frequencies. We show the amplitude spectrum for Sector\,48\,--\,51 data in Fig.\,\ref{fig:USC_ft4}, and list the prewhitened frequencies in Table\,\ref{tab:USCfreq}. The amplitude spectrum looks quite interesting. It contains two high amplitude frequencies, f$_2$ and f$_5$. The higher one, f$_5$, is surrounded by two nearby frequencies spaced by 0.67\,\uHz, on average. If this is the rotationally split triplet, the rotation period is 17.3\,d. In addition, both frequencies f$_2$ and f$_5$ are surrounded by low amplitude frequencies spaced exactly by 2.09\,\uHz. This frequency spacing would give 5.5\,d rotation period. However, we lean towards interpreting the latter period as an orbital period. We do not detect any signal at 2.09\,\uHz, which would confirm our hunch, so follow-up spectroscopic velocities would be required to determine if this star is a binary.

\paragraph{TIC\,207440586}
(LM\,Dra) is a known sdOB pulsator. It was classified as an sdB by \citet{green86}. Our fit to a spectrum taken with the NOT in 2022 gives \teff\,=\,34\,037(132)\,K, \loggcms\,=\,5.72(3) and \logy\,=\,-1.58(7), and we classified the star as an sdOB. \citet{silvotti00} reported a detection of two pulsation periods. \citet{reed07b} analyzed an extended coverage photometric data and reported six frequencies. \tess\ observed the star during Sectors\,49\,--\,52 and 56. The pulsations are very unstable in this star and merging all data together makes prewhitening very challenging. We decided to analyze only Sector\,56 since they are described with a significantly lower noise level and allow for detection of all signals that are present in each sector data set but with higher S/N. We detected three frequencies, all of them previously reported by \citet{reed07b}. We show the amplitude spectrum for Sector\,56 data in Fig.\,\ref{fig:USC_ft4}, and list the prewhitened frequencies in Table\,\ref{tab:USCfreq}.

\paragraph{TIC\,219492314}
(V1078\,Her) is a known sdOB pulsator. It was classified as an sdOA by \citet{green86}, while as an sdO by \citet{wegner93}. \citet{lei19a} identified the star as an sdOB. \citet{kuassivi05} detected one pulsation mode at the surface. \tess\ observed the star during Sectors\,51 and 52. We detected only one frequency, which is the same one as reported by \citet{kuassivi05}. We show the amplitude spectrum in Fig.\,\ref{fig:USC_ft4}, and list the prewhitened frequencies in Table\,\ref{tab:USCfreq}.

\paragraph{TIC\,240868270}
(GD\,274) is a new sdOB pulsator. For a long time it was considered to be a white dwarf candidate until \citet{lei23} classified the star as an sdOB+F9.5VI system. \tess\ observed the star during Sector\,58. We detect five high frequencies associated with stellar pulsations and three low frequencies likely related to the binary flux variation. The latter variation is highly non-sinusoidal and three frequencies are not linear combinations, including harmonics. We denoted them with upper case letters. We show the amplitude spectrum  in Fig.\,\ref{fig:USC_ft4}, and list the prewhitened frequencies in Table\,\ref{tab:USCfreq}.

\paragraph{TIC\,266013993}
(PG\,0048+091) is a known sdB pulsator. \citet{koen04} classified the star as an sdB and reported a flux variability interpreted as stellar pulsations. \citet{reed07b} listed 28 frequencies noting that the amplitude spectrum is very variable. PG\,0048+091 was observed as part of the \emph{K2} mission \citep{reed19} and we refer the reader to that paper as the most complete study. \tess\ observed it during Sectors\,42 and 43 and we detect four frequencies with well defined profiles indicating they are amplitude/frequency stable during \tess's two month coverage. We find frequencies in the list presented by \citet{reed07b} within a few \uHz. We show the amplitude spectrum in Fig.\,\ref{fig:USC_ft4}, and list the prewhitened frequencies in Table\,\ref{tab:USCfreq}.

\paragraph{TIC\,273255412}
(GALEX\,J201337.6+092801) is a known sdB pulsator. It was classified as an sdB and a flux variability interpreted as stellar pulsations was discovered by \citet{ostensen11}. The authors listed 16 frequencies. \tess\ observed the star during Sector\,54. We detect 26 frequencies, however the profiles of peaks are complex, which means that the frequencies/amplitudes are not stable or there are unresolved multiplets. Compared with the list presented by \citet{ostensen11} we confirm eight frequencies, four are within a few \uHz, while another four were not detected before. Frequency f$_1$ was not detected before and if it is a pulsation frequency it can be associated with gravity modes, which would make the star a hybrid pulsator. We show the amplitude spectrum in Fig.\,\ref{fig:USC_ft4}, and list the prewhitened frequencies in Table\,\ref{tab:USCfreq}.

\paragraph{TIC\,284692897}
(V2214\,Cyg, KPD\,1930+2752) is a known sdB pulsator. It was classified as an sdB by \citet{downes86}. \citet{billeres00} reported 44 pulsation frequencies and an ellipsoidal variation in the sdB+WD binary system. \citet{maxted00} concluded that the star is a good candidate for the progenitor of a Type Ia supernova of this type, which will merge on an astrophysically interesting timescale. \citet{reed11} analyzed multisite photometric data and reported 68 frequencies and additional 13 suspected frequencies. They also reported tidally-tipped pulsations as well as binary-phase-dependent pulsations produced by tidal forces. Our fit to a spectrum taken with the NOT in 2022 gives \teff\,=\,34\,594(236)\,K, \loggcms\,=\,5.53(5) and \logy\,=\,-1.61(8), and we classified the star as an sdB. \tess\ observed the star during Sectors\,40, 41 and 54. We merged and analyzed only the consecutive Sectors\,40 and 41 data. We detect 14 pulsation frequencies and two related to binarity. For an ellipsoidal variable the dominant amplitude frequency is half the orbital period. All but four pulsation frequencies were reported by \citet{reed11}. The new frequencies are f$_5$, f$_7$, f$_{13}$ and f$_{14}$. We show the amplitude spectrum for Sectors\,40 and 41 in Fig.\,\ref{fig:USC_ft5}, and list the prewhitened frequencies in Table\,\ref{tab:USCfreq}.

\paragraph{TIC\,309807601}
(PG\,1455+501) is a new sdB pulsator. It was classified as an sdB by \citet{green86}, while \citet{geier17} reported an sdB+F2 classification. \tess\ observed the star during Sectors\,49 and 50. We detect two high frequencies. We show the amplitude spectrum in Fig.\,\ref{fig:USC_ft5}, and list the prewhitened frequencies in Table\,\ref{tab:USCfreq}.

\paragraph{TIC\,310937915}
(EP\,Psc) is a known sdOB pulsator. It was classified as an sdB by \citet{green86}. \citet{silvotti02a} derived spectroscopic parameters that indicate an sdOB type and they reported the discovery of three pulsation periods. Our fit to the average of 10 spectra taken with the NOT gives \teff\,=\,34\,465(131)\,K, \loggcms\,=\,5.74(2) and \logy\,=\,-1.67(5). \tess\ observed the star during Sector\,42. We detected only one frequency, which was the highest amplitude one reported by \citet{silvotti02a}. We show the amplitude spectrum in Fig.\,\ref{fig:USC_ft5}, and list the prewhitened frequency in Table\,\ref{tab:USCfreq}.

\paragraph{TIC\,355754830}
(GALEX\,J063952.0+515658) is a known sdB pulsator. It was classified as an sdB by \citet{nemeth12}. \citet{vuckovic12} reported the results of an analysis of multi-site photometric data and reported a signal in both high and low frequency regions, making it a hybrid pulsator. \tess\ observed the star during Sector\,60. We detect 14 frequencies, four and 10, in the low and high frequency regions, respectively, confirming it as a hybrid pulsator. We stress that some of the frequencies around the highest amplitude frequency may be residuals of amplitude/phase instability and not independent frequencies. We show the amplitude spectrum in Fig.\,\ref{fig:USC_ft5}, and list the prewhitened frequencies in Table\,\ref{tab:USCfreq}.

\paragraph{TIC\,357232133}
(GALEX\,J074149.0+552451) is a new sdOB pulsator. It was classified as an sdB by \citet{nemeth12} and reclassified as an sdOB by \citet{geier17}. \tess\ observed the star during Sectors\,47 and 60. To avoid a complex window function we analyzed Sector\,47 data only. We detect four frequencies. We show the amplitude spectrum for Sector\,47 data in Fig.\,\ref{fig:USC_ft5}, and list the prewhitened frequencies in Table\,\ref{tab:USCfreq}.

\paragraph{TIC\,364966239}
(\gaia\,DR3 2235556213015091456) is a new sdOB pulsator. Our fit to a spectrum taken with the NOT in 2019 gives \teff\,=\,33\,587(241)\,K, \loggcms\,=\,5.70(5) and \logy\,=\,-1.50(7), and we classified the star as an sdOB. \tess\ observed the star during Sectors\,41, 47, 54\,--\,57 and 60. To avoid complex peak profiles and difficulties with prewhitening, we combined and analyzed Sectors\,54\,--\,57 and 60. We detect four frequencies, with three of them indicating a rotationally split triplet. The average frequency splitting indicates a rotation period of 20.141(21)\,d. We show the amplitude spectrum for Sectors\,54\,--\,57 and 60 data in Fig.\,\ref{fig:USC_ft5}, and  list the prewhitened frequencies in Table\,\ref{tab:USCfreq}.

\paragraph{TIC\,397595169}
(V391\,Peg) is a known sdB pulsator. It was found to be a pulsator and classified as an sdB by \citet{ostensen01b}. The authors reported one high frequency. \citet{silvotti02b} analyzed multi-site photometric data and detected five frequencies. Pulsation frequencies were used by \citet{silvotti07} to infer the presence of a planet orbiting the sdB star, however its presence is now uncertain \citep{silvotti18}. \tess\ observed the star during Sector\,56. We detect two frequencies, which were found to be the two highest in \citet{silvotti02b}. We show the amplitude spectrum in Fig.\,\ref{fig:USC_ft5}, and list the prewhitened frequencies in Table\,\ref{tab:USCfreq}.

\paragraph{TIC\,424720852}
(GALEX\,J193832.5+560944) is a known sdB pulsator. \citet{holdsworth17} derived spectroscopic parameters and classified the star as an sdB. They detected one short period and interpreted it as a stellar pulsation. \tess\ observed the star during Sectors\,40, 41, 50, 54\,--\,57 and 60. We split all data into two subsets that we analyzed separately. The first one includes the first two sectors, while the second one contains the remaining data sectors. We detect signals in both low and high frequency regions. In the former we found one and five frequencies in the first and second subsets, respectively. In the latter we found eight frequencies in both subsets. We show the amplitude spectrum calculated from Sectors\,50, 54\,--\,57 and 60, in Fig.\,\ref{fig:USC_ft6}, and list the prewhitened frequencies in Table\,\ref{tab:USCfreq}.

\paragraph{TIC\,441725813}
(\gaia\,DR3 1655107708129775744) is a new sdB pulsator. Our fit to a spectrum taken with the NOT in 2017 gives \teff\,=\,27\,204(141)\,K, \loggcms\,=\,5.37(2) and \logy\,=\,-2.93(5), and we classified the star as an sdB. \tess\ observed the star during Sectors\,40, 41, 47\,--\,52 and 55\,--\,60. The amplitude spectrum is very rich in frequencies in the low frequency region. It makes the star a g-mode-dominated sdB pulsator. To avoid a complex window function we separated all data into three subsets to make them continuous in coverage. The most signals exist in the subset of Sectors\,55\,--\,60 data. The other subsets do not show any additional signals. We are aware that merging all data together would lower the noise level in an amplitude spectrum the most, however unstable pulsation modes contribute with residuals too strongly and no extra frequencies around the dominant signals can be conveniently detected. In addition, we only prewhitened signals above 2300\,\uHz, which can be associated with pressure modes. Frequencies in the low frequency region need individual analysis, which is beyond the scope of this work. We detect 14 frequencies that we associated with pressure modes. The most noticeable signal appears around 2769\,\uHz. There are three frequencies symmetrically spaced, which we interpret as rotationally split modes, suggesting a rotation period of 17.188(18)\,d. We show the amplitude spectrum for Sectors\,55\,--\,60 data in Fig.\,\ref{fig:USC_ft6}, and list the prewhitened frequencies in Table\,\ref{tab:USCfreq}.
%An in-depth study of this remarkable sdB pulsator is provided by Su et al. (submitted).

\paragraph{TIC\,471015194}
(\gaia\,DR3 2238354611842566912) is a known sdB pulsator. \cite{prins19} reported a discovery of a frequency at 5789.6\,\uHz. Our fit to a spectrum taken with the NOT in 2017 gives \teff\,=\,33\,150(592)\,K, \loggcms\,=\,5.71(12) and \logy\,=\,-2.93(44), and we classified the star as an sdB. \tess\ observed the star during Sectors\,40, 41, 47, 50, 54\,--\,57 and 60 in which we detect five frequencies. The highest amplitude frequency, f$_3$, overlaps with the one reported by \citet{prins19}. We show the amplitude spectrum in Fig.\,\ref{fig:USC_ft6}, and list the prewhitened frequencies in Table\,\ref{tab:USCfreq}.

\section{Discussion}
In the Northern Ecliptic hemisphere we found 35 sdB and 15 sdOB p-mode pulsating stars, including 27 (18 sdB and nine sdOB) new pulsators. Overall, including the Southern Ecliptic hemisphere, with \tess\ data we detect p-mode pulsations in 93 hot subdwarfs, including 67 sdB, 23 sdOB, two sdO and one He-sdOB stars. Among the 48 new pulsators discovered with \tess, we identified 35 sdB, 12 sdOB and one He-sdOB stars.

\subsection{Detection of pulsating hot subdwarfs with \tess}
As this paper completes a \tess\ survey of both hemispheres, it is prudent to do a completeness study. However, such a study has many complexities, including but not limited to that \tess\ is a small-aperture telescope with large pixels and sdB stars tend to be on the faint end of its detection capability. Pulsation amplitudes in sdBV stars also vary substantially between stars, and often even within the same star over time \citep[e.g.][]{kilk2010}. The highest pulsation amplitude of stars in this paper varies from under 1 to over 70 ppt, greatly affecting how bright the star would need to be to have sufficient signal-to-noise to be detected by \tess. Additionally, some stars are observed during multiple sectors, which reduces the detection threshold. To do a proper, statistically-significant determination would require an examination of all non-pulsators as well as the pulsators in this paper and likely should also include the g-mode pulsators which we exclude. Pixel decontamination from nearby stars and multi-sector normalization would also need to be completed. That analysis will be a future paper unto itself, but we will do a  first-cut simplification of detection likelihood.

\begin{figure}
\centering
\includegraphics[width=\hsize]{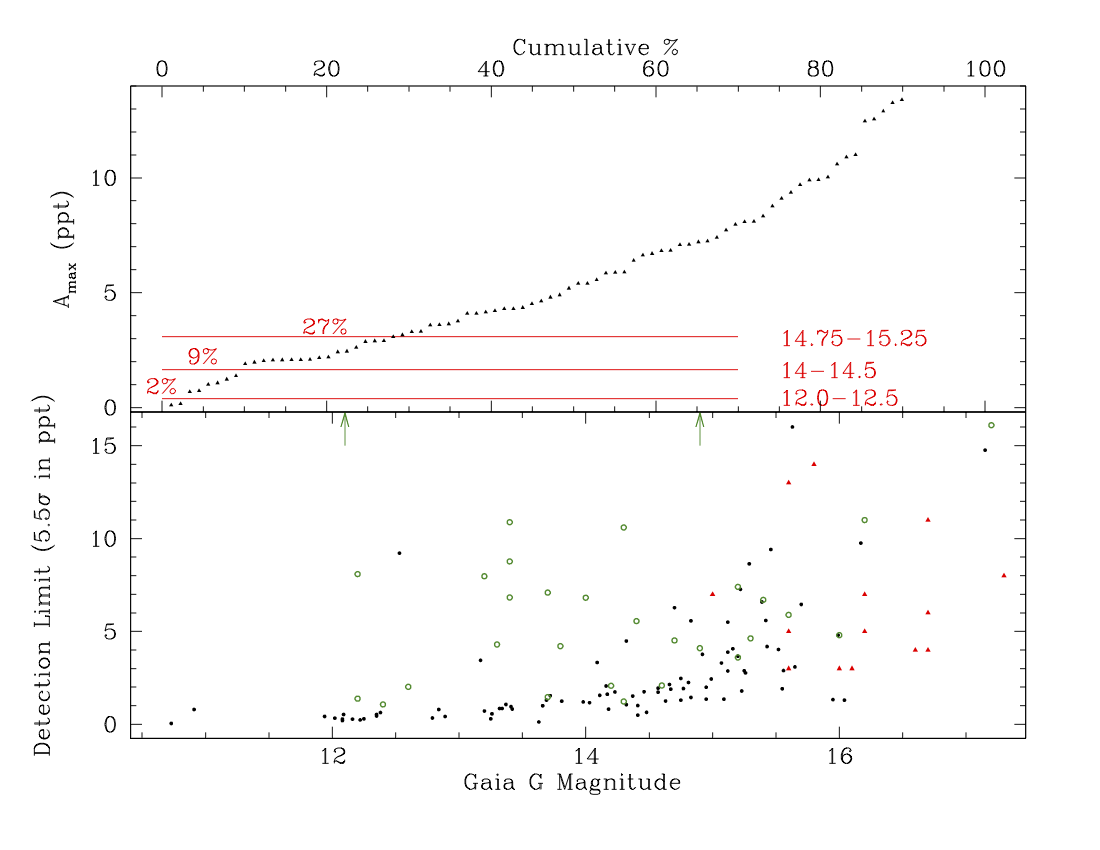}
\caption{Tests of detection completeness. Bottom panel: Black circles indicate single-sector detection limits of p-mode sdBV stars observed by \tess. Green circles and red triangles are not detection limits, but A$_{\rm max}$ for stars listed in \citet{ostensen10a} that were observed by \tess. The green circles indicate stars in which pulsations were detected (amplitudes from \tess\ data), while red triangles indicate stars in which pulsations were not detected by \tess, \citet[amplitudes as listed in][]{ostensen10a}. Green arrows indicate the stars TIC\,90960668 and TIC\,88565376, which have amplitudes too high to appear in this figure. Top panel: A$_{\rm max}$ of p-mode sdBV stars (black triangles) detected by \tess\ from low to high amplitude. The red lines indicate detection limits for those magnitude ranges and what fraction of detected sdBV stars would be undetected based on A$_{\rm max}$.}
\label{fig:det_lim}
\end{figure}

This is done two ways. The first is to compare pulsation detection thresholds with brightness to the pulsation amplitudes actually observed by \tess. As pulsation amplitudes are brightness independent, the pulsations listed in these papers should be representative of p mode sdBV pulsations in general. By comparing the highest-amplitude pulsation for each star (A$_{\rm max}$, as only one pulsation needs to be detected to be considered a pulsator) with magnitude-averaged detection limits, we can estimate what fraction of pulsators would likely not be detected for a given magnitude. The black dots in the bottom panel of Fig.\,\ref{fig:det_lim} are the 5.5$\sigma$ detection limits for each of the \tess-observed p-mode pulsators. As expected, thresholds increase for fainter stars and beyond G=15.25, the scatter is quite large. We calculate average detection limits for the ranges 12-12.5 (0.385 ppt), 14-14.5 (1.65 ppt), and 14.75-15.25 (3.08 ppt) and in the top panel of Fig.\,\ref{fig:det_lim} we compare that with A$_{\rm max}$ of each \tess-observed p-mode pulsator, accumulating from lowest to highest amplitude. To keep the lower amplitudes visible, we cut off at the 90th percentile (14\,ppt). Red horizontal lines indicate the average detection limit for the three brightness ranges from the bottom panel and amplitudes below those lines indicate stars which would not likely be detected as pulsators for that brightness limit. For single-sector \tess\ observations, the lowest-amplitude 2\% of our pulsators would have been missed at magnitude 12-12.5, 9\% at 14-14.5, and 27\% at 14.75-15.25.

A second simple exercise is to examine a set of p-mode pulsators known previous to \tess\ that were observed by \tess\ to see if we found them as pulsators. For this exercise we used the 49 sdBV stars listed in Table 9 of \citet{ostensen10a} of which \tess\ observed 45. \tess\ detected 30 as pulsators, with the \tess\  A$_{\rm max}$ for those stars indicated with green open circles in the bottom panel of Fig.\,\ref{fig:det_lim}. \tess\ did not detect 14 stars and their A$_{\rm max}$, as listed in \citet{ostensen10a}, are indicated with red triangles. 24 of 25 (96\%) stars brighter than magnitude 15.25 were detected by \tess\, but only 6 of 19 (32\%) fainter stars were detected. The undetected star PG\,1419+081, with V=14.9, has an A$_{\rm max}$ of 7\,ppt in \citet{ostensen10a} and we find a \tess\ detection threshold of 3.00\,ppt, so clearly it should have been detected were the pulsation amplitudes the same as in \citet{ostensen10a}.

The results from the first of these simplistic tests indicate that \tess\ likely missed some bright (G<15.25) p-mode sdB pulsators whereas the second test suggests incomplete detections only for G>15.25. However, the ground-based \citet{ostensen10a} sample, which was observed for much shorter durations, all had A$_{\rm max}\geq$\,3\,ppt. This is likely a selection effect of the shorter-duration observations whereas our month-long TESS observations detect pulsators with A$_{\rm max}$ as low as 1\,ppt. For both tests, beyond G=15.25, \tess\ very likely misses pulsators and we would suggest that even some bright ones were missed. A more rigorous detection likelihood will be completed in an upcoming paper.

\begin{figure}
\centering
\includegraphics[width=\hsize]{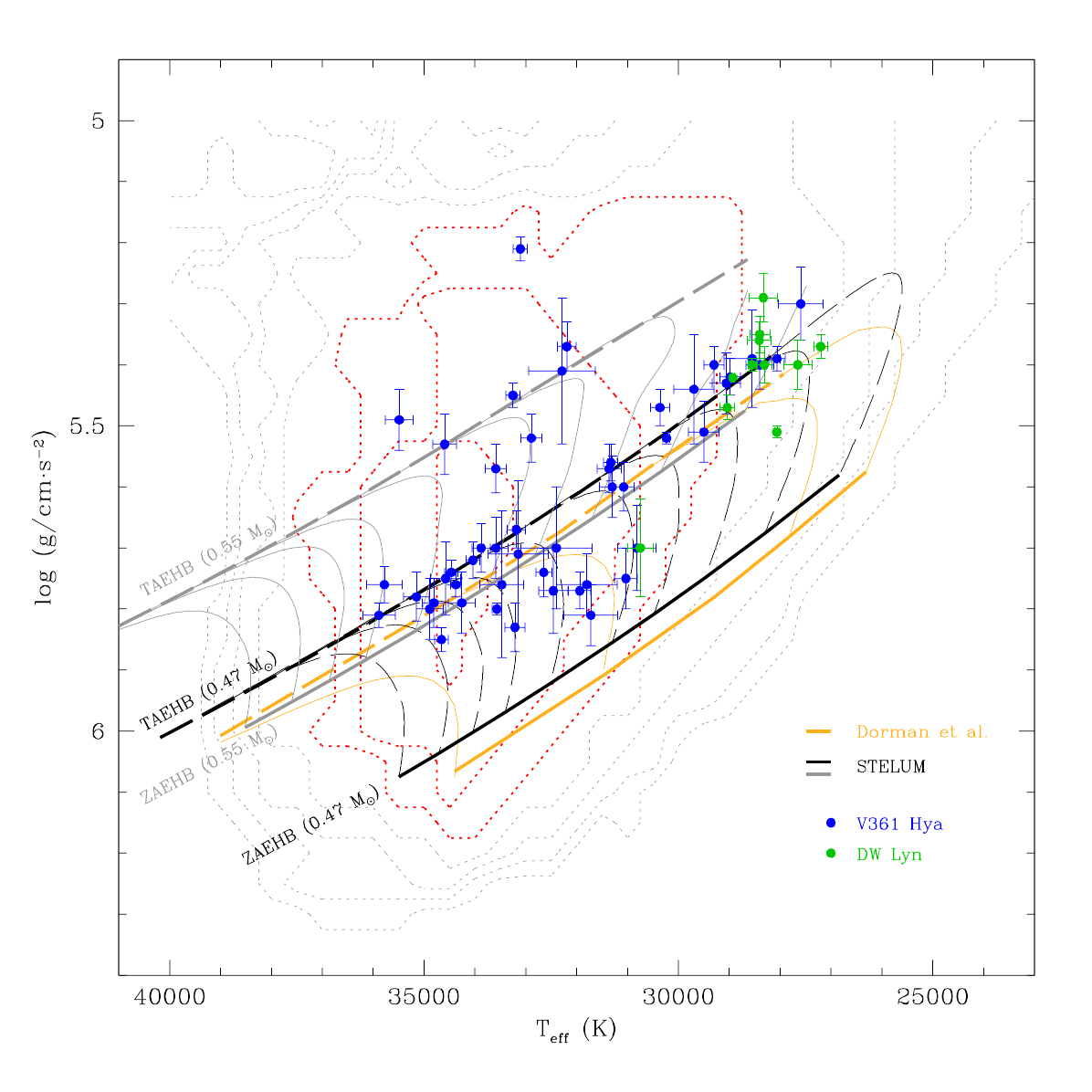}
\caption{Distribution, in the surface gravity (\logg)\,--\,effective temperature (\teff) plane, of the hot subdwarf stars observed by TESS throughout cycle 1 to 4 and showing p-mode pulsations. Pure p-mode pulsators (V361\,Hya stars) are represented as blue dots and hybrid pulsators featuring also g-modes (DW\,Lyn stars) are shown as green dots. Superimposed to this sample, the ZAEHB and TAEHB positions are indicated for an assumed core mass of 0.47\,M$_\odot$ (black curves) and 0.55\,M$_\odot$ (grey curves), illustrating how the EHB location depends on stellar mass. The models that materialize the EHB tracks are computed with the STELUM code for the specified core mass and varying envelope masses such as D(H)$\equiv \log$[M$_{\rm env}$/(M$_{\rm core}$+M$_{\rm env}$)]\,=\,-6.0, -5.0, -4.5, -4.0, -3.5, -3.0, -2.75, -2.5, -2.25, -2.0, from high to low \teff. These models incorporate in their envelope nonuniform iron abundance profiles derived from equilibrium between radiative levitation and gravitational settling (see text for details).
For comparison purposes, typical 0.47\,M$_\odot$ tracks assuming solar uniform envelope composition are also represented (yellow curves; \citealt{charpinet00,dorman93}). 
Dotted-line contours in the background represent the predicted number of excited $\ell=0$ $p$-modes from non-adiabatic pulsation calculations as a proxy of the $\kappa$-mechanism efficiency \citep{charpinet01}. The three innermost contours (highlighted in red) materialize the region of the \logg\,--\,\teff\ plane where the driving of p-modes is the strongest.
}
\label{fig:teff_logg}
\end{figure}

\subsection{Evolutionary status and instability strip}
Fig.\,\ref{fig:teff_logg} shows the distribution, in the \logg\,--\,\teff\ plane, of p-mode and hybrid hot subdwarf pulsators monitored by \tess\ in both the Southern and Northern Ecliptic hemispheres. Only stars with reliable estimates of their atmospheric parameters from spectroscopy are represented (see Table\,\ref{tab:spec_fit}). 
This sample allows us to more densely map the instability region where p-mode pulsations are found and compare with evolutionary tracks and theoretical predictions from non-adiabatic pulsation calculations.

In this diagram, the Zero Age and Terminal Age Extreme Horizontal Branches (ZAEHB and TAEHB) are determined from a set of static hot subdwarf models calculated with the Montr\'eal stellar structure code STELUM \citep{vangrootel13}. These models span the appropriate parameter range in terms of central helium content and H-rich envelope mass to cover the entire core helium burning phase (CHeB) and the relevant part of the EHB where p-mode pulsators are located. This is unlike many diagrams of the same kind found in the literature showing tracks from standard evolutionary models. Here, we exploit the greater flexibility of static structures to illustrate various factors that may affect the position of the theoretical EHB and that are often overlooked. 

\begin{table*}
\centering
\caption{Spectroscopic estimates of pulsating hot subdwarfs reported here and in Paper I, for which we derived an acceptable spectroscopic fits. Archive spectra were (re-)fit using the same model and method as described in Section\,\ref{sec:spectra}.}
\label{tab:spec_fit}
\begin{tabular}{rrrrrrrrrrr}
\hline
\multicolumn{1}{c}{TIC} & \multicolumn{1}{c}{\teff} & \multicolumn{1}{c}{$\sigma$} & \multicolumn{1}{c}{\loggcms} & \multicolumn{1}{c}{$\sigma$} & \multicolumn{1}{c}{\logy} & \multicolumn{1}{c}{$\sigma$} & \multicolumn{1}{c}{Telescope} & \multicolumn{1}{c}{Reference} \\%& \multicolumn{1}{c}{J\,name} & \multicolumn{1}{c}{Name} \\
%\multicolumn{1}{c}{TIC} & \multicolumn{1}{c}{\teff} & \multicolumn{1}{c}{$\sigma_{\teff}$} & \multicolumn{1}{c}{\loggcms} & \multicolumn{1}{c}{$\sigma_{\logg}$} & \multicolumn{1}{c}{\logy} & \multicolumn{1}{c}{$\sigma_{\logy}$} \\
\hline
    968226   &  28\,982 &  198  &   5.42  &  0.03   &  -3.15  &  0.09    &   NOT      &                            \\  %%  & J21552-1025  & PHL 211                        
   4632676   &  33\,589 &  210  &   5.57  &  0.04   &  -1.64  &  0.05    &   SDSS     &  archive                   \\  %%  & J12036+2531  & Ton 74                         
  16993518   &  32\,194 &  187  &   5.37  &  0.04   &  -2.27  &  0.09    &   NOT      &                            \\  %%  & J15409+3518  & FBS 1539+355                   
  19690565   &  31\,033 &  225  &   5.75  &  0.05   &  -3.12  &  0.15    &   NOT      &  \citet{baran23}           \\  %%  & J08212-0828  & Gaia DR2 5753155495252812544   
  26291471   &  34\,376 &  106  &   5.76  &  0.02   &  -1.34  &  0.03    &   VLT      &  archive                   \\  %%  & J14534-1009  & HE 1450-0957                   
  33318760   &  34\,806 &  233  &   5.79  &  0.04   &  -1.68  &  0.06    &   NTT      &  \citet{randall09}       \\  %%  & J10006-1151  & V541 Hya                       
  47377536   &  35\,149 &  251  &   5.78  &  0.04   &  -1.84  &  0.06    &   LAMOST   &  archive                   \\  %%  & J10500-0000  & UY Sex                         
  55753808   &  34\,889 &  263  &   5.80  &  0.05   &  -1.80  &  0.06    &   LAMOST   &  archive                   \\  %%  & J12064+5709  & PG 1203+574                    
  56863037   &  31\,940 &  220  &   5.77  &  0.03   &  -2.55  &  0.06    &   2.3mKP   &  \citet{randall07}       \\  %%  & J09149+4523  & DT Lyn                         
  62381958   &  35\,779 &  352  &   5.76  &  0.03   &  -1.30  &  0.00    &   NTT      &  archive                   \\  %%  & J01565-1354  & EC 01541-1409                  
  63168679   &  35\,885 &  318  &   5.81  &  0.02   &  -1.60  &  0.03    &   LAMOST   &  archive                   \\  %%  & J07534+2324  & Gaia DR2 675213084211549696    
  68495594   &  33\,253 &  137  &   5.45  &  0.02   &  -1.40  &  0.05    &   NOT      &                            \\  %%  & J06534+3303  & HD 265435                      
  69298924   &  28\,063 &  163  &   5.39  &  0.02   &  -2.97  &  0.06    &   NOT      &  \citet{baran11}         \\  %%  & J08069+1527  & GALEX J080656.7+152718         
  70549283   &  33\,879 &  178  &   5.70  &  0.04   &  -1.51  &  0.05    &   LAMOST   &  archive                   \\  %%  & J07403+2049  & SDSS J074023.56+204937.0       
  85145647   &  32\,287 &  656  &   5.41  &  0.12   &  -3.03  &  0.39    &   INT      &                            \\  %%  & J07304+6211  & FBS 0725+623                   
  88484868   &  29\,040 &  140  &   5.47  &  0.02   &  -2.85  &  0.06    &   INT      &  archive                   \\  %%  & J07035+6236  & FBS 0658+627                   
  88565376   &  28\,400 &  200  &   5.35  &  0.03   &  -2.7   &  0.03    &   3.5mCA   &  \citet{dreizler02}      \\  %%  & J07071+6038  & DW Lyn                              
  90960668   &  28\,932 &   23  &   5.421 &  0.003  &  -2.721 &  0.003   &   NOT      &  \citet{ostensen07}      \\  %%  & J23153+2905  & V585 Peg                        
  95752908   &  34\,656 &  133  &   5.85  &  0.02   &  -1.31  &  0.03    &   NOT      & \citet{holdsworth17}    \\  %%  & J09020-0720  & TYC 4890-19-1                    
 114196505   &  27\,594 &  442  &   5.30  &  0.06   &  -2.99  &  0.19    &   NOT      &                            \\  %%  & J19240-1256  &                                
 136975077   &  31\,800 &  600  &   5.76  &  0.05   &  -2.23  &  0.1     &   Keck     &  \citet{heber00}         \\  %%  & J21117+4413  & V2203 Cyg                      
 137502282   &  34\,259 &  267  &   5.79  &  0.05   &  -1.31  &  0.07    &   INT      &                            \\  %%  & J08518+7901  & FBS 0844+792                   
 138618727   &  29\,500 &  300  &   5.51  &  0.05   &  -2.93  &  0.05    &   Keck     &  \citet{heber00}         \\  %%  & J11472+6115  & KL UMa                         
 139481265   &  31\,720 &  529  &   5.81  &  0.05   &  -2.23  &  0.08    &   NOT      &  \citet{baran23}           \\  %%  & J05529+1118  & Gaia DR2 3342874205845523072   
 142200764   &  30\,235 &   66  &   5.52  &  0.01   &  -2.84  &  0.04    &   VLT      &  archive                   \\  %%  & J02329-4310  & HE 0230-4323                    
 154818961   &  28\,330 &  281  &   5.29  &  0.04   &  -1.71  &  0.05    &   NOT      &                            \\  %%  & J22581+4614  & Gaia DR3 1935962732084366592   
 156618553   &  28\,065 &   43  &   5.51  &  0.01   &  -2.45  &  0.02    &   VLT      &  \citet{vuckovic14}      \\  %%  & J12443-0840  & HW Vir                         
 157141133   &  31\,076 &  207  &   5.60  &  0.04   &  -2.97  &  0.16    &   INT      &                            \\  %%  & J20247-1022  & GALEX J20247-1022              
 159644241   &  32\,035 &  117  &   5.66  &  0.02   &  -2.13  &  0.04    &   NOT      &  \citet{baran12}         \\  %%  & J19249+4707  & Saradoc                        
 165312944   &  33\,570 &   88  &   5.80  &  0.01   &  -1.47  &  0.03    &   NOT      &  \citet{reed09}          \\  %%  & J12214+5304  & KY UMa                         
 175402069   &  31\,300 &  250  &   5.60  &  0.05   &  -2.93  &  0.05    &   VLT      &  \citet{vuckovic07}      \\  %%  & J13388-0201  & NY Vir                          
 178081355   &  33\,186 &  176  &   5.67  &  0.03   &  -1.77  &  0.03    &   SDSS     &  archive                   \\  %%  & J03183+4155  & FBS 0315+417                    
 186484490   &  28\,401 &  274  &   5.40  &  0.04   &  -2.90  &  0.09    &   LAMOST   &  archive                   \\  %%  & J01203+3950  & FBS 0117+396                    
 191442416   &  32\,400 &  700  &   5.70  &  0.10   &  -2.20  &  0.20    &   2.2mCA   &  \citet{ostensen01a}     \\  %%  & J00427+4319  & V429 And                          
 199715319   &  28\,144 &  310  &   5.32  &  0.04   &  -3.15  &  0.10    &   NOT      &                            \\  %%  & J16578+5511  & PG 1656+553                    
 202354658   &  29\,060 &  280  &   5.43  &  0.05   &  -3.44  &  0.26    &   NTT/4mKP &  \citet{nemeth12}        \\  %%  & J15451+5955  & PG 1544+601                    
 207440586   &  34\,037 &  132  &   5.72  &  0.03   &  -1.58  &  0.07    &   NOT      &                            \\  %%  & J16194+5606  & LM Dra                         
 219492314   &  34\,575 &  299  &   5.75  &  0.06   &  -1.34  &  0.08    &   LAMOST   &  archive                   \\  %%  & J16147+4227  & V1078 Her                       
 222892604   &  30\,363 &  188  &   5.47  &  0.03   &  -3.08  &  0.13    &   NOT      &                            \\  %%  & J19013+0614  & LS IV +06 5                    
 248776104   &  32\,891 &  205  &   5.52  &  0.04   &  -2.89  &  0.14    &   NOT      &                            \\  %%  & J20555-0004  & PG 2052-003                    
 266013993   &  32\,460 &  290  &   5.77  &  0.07   &  -2.33  &  0.11    &   NOT      &  \citet{reed09}          \\  %%  & J00514+0921  & PG 0048+091                    
 273255412   &  33\,108 &  140  &   5.21  &  0.02   &  -2.63  &  0.07    &   LAMOST   &  archive                   \\  %%  & J20136+0928  & GALEX J20136+0928              
 284692897   &  34\,594 &  236  &   5.53  &  0.05   &  -1.61  &  0.08    &   NOT      &                            \\  %%  & J19322+2758  & KPD 1930+2752                  
 310937915   &  34\,465 &  131  &   5.74  &  0.02   &  -1.67  &  0.05    &   NOT      &                            \\  %%  & J23063-0209  & EP Psc                         
 331656308   &  28\,411 &  230  &   5.36  &  0.03   &  -2.95  &  0.11    &   NOT      &                            \\  %%  & J04406+7603  & FBS 0433+759                   
 355754830   &  30\,750 &  250  &   5.70  &  0.08   &  -2.59  &  0.29    &   4mKP     &  \cite{vennes11}        \\  %%  & J06398+5157  & GALEX J063952.0+515658           
 357232133   &  33\,470 &  430  &   5.76  &  0.12   &  -1.93  &  0.17    &   NTT/4mKP &  \citet{nemeth12}        \\  %%  & J07418+5524  & GALEX J074149.0+552451         
 364966239   &  33\,587 &  241  &   5.70  &  0.05   &  -1.50  &  0.07    &   NOT      &                            \\  %%  & J19528+5808  & Gaia DR2 2235556213015091456   
 366656123   &  31\,331 &  140  &   5.56  &  0.03   &  -2.82  &  0.06    &   SDSS     &  \citet{baran23}           \\  %%  & J08413+0630  & SDSS J084122.67+063029.6       
 392092589   &  27\,654 &  280  &   5.40  &  0.04   &  -2.97  &  0.10    &   NOT      &                            \\  %%  & J08014+2557  & SDSS J080127.57+255742.9       
 396954061   &  33\,214 &  199  &   5.83  &  0.04   &  -1.57  &  0.04    &   NOT      &                            \\  %%  & J04158+0154  & UCAC232311775                  
 397595169   &  29\,300 &  200  &   5.40  &  0.03   &  -3.00  &  0.15    &   3.5mCA   &  \citet{ostensen01b}     \\  %%  & J22042+2625  & V391 Peg                         
 424720852   &  28\,311 &  149  &   5.40  &  0.03   &  -2.79  &  0.07    &   INT      &  archive                   \\  %%  & J19385+5609  & 2MASSJ19383247+5609446          
 436579904   &  31\,360 &  240  &   5.57  &  0.04   &  -2.65  &  0.10    &   NOT      &  \citet{reed20}          \\  %%  & J04449+1416  & V1405 Ori                      
 437043466   &  28\,557 &   82  &   5.40  &  0.01   &  -3.07  &  0.02    &   NOT      &  \citet{baran17}         \\  %%  & J08568+1701  & 2M0856+1701                    
 441725813   &  27\,204 &  141  &   5.37  &  0.02   &  -2.93  &  0.05    &   NOT      &                            \\  %%  & J17049+7304  & 2M1704+7304                    
 455095580   &  30\,817 &  380  &   5.70  &  0.07   &  -2.60  &  0.16    &   NOT      &  \citet{baran23}           \\  %%  & J08127+0600  & SDSS J081243.64+060030.5       
 471013461   &  28\,551 &  548  &   5.39  &  0.08   &  -2.94  &  0.02    &   SOAR     &  archive                   \\  %%  & J03547-4015  & Gaia DR2 4843383737223292672
 471015194   &  33\,150 &  592  &   5.71  &  0.12   &  -2.93  &  0.44    &   NOT      &                            \\  %%  & J19384+5824  & PSO J294.6167+58.4043
 673345538   &  29\,693 &  396  &   5.44  &  0.09   &  -2.72  &  0.21    &   LAMOST   &  archive                   \\  %%  & J04552+1305  & RAT J0455+1305 
\hline
\end{tabular}
\end{table*}

A first factor is related to the actual composition of a hot subdwarf envelope which is known to be affected by microscopic diffusion, while standard evolution models usually assume a solar homogeneous composition. Diffusion has an impact on the EHB location because the sedimentation of helium and distribution of opaque elements (in particular those from the iron group) significantly change the thermal structure of the stellar envelope and consequently the surface parameters for a given envelope mass. The trend can be seen in Fig.\,\ref{fig:teff_logg} by comparing the 0.47\,M$_\odot$ tracks derived from the STELUM models, which incorporate a double layered H/He envelope with nonuniform iron-abundance distributions from calculations of equilibrium between radiative levitation and gravitational settling (see \citealt{charpinet97} and \citealt{vangrootel13} for details), with standard evolution models of solar composition from \citet{dorman93}. The difference between the two is a noticeable shift of the ZAEHB and TAEHB towards lower surface gravities when microscopic diffusion is included, actually providing a better match to the distribution of observed hot subdwarf stars. 

A second factor that is generally overlooked is the influence of stellar mass. Most stars that form the EHB are expected to have the canonical mass of $\sim$0.47\,M$_\odot$, corresponding to the critical value for helium ignition in degenerate conditions (the so-called helium flash), since their progenitors on the main-sequence are generally believed to be less than 2\,M$_\odot$ stars evolving through the red giant phase. This expectation is indeed mostly verified as the observed EHB appears as a rather homogeneous group that can be reasonably well reproduced, globally, with models assuming the same core mass. Nevertheless, even if the mass distribution of hot subdwarf stars is strongly peaked at the canonical value \citep{fontaine12}, scatter around that value is still expected that will lead to some fuzziness in the position of the ZAEHB and TAEHB. In addition, some binary evolution channels that are also invoked in the formation of hot subdwarfs -- for instance, those involving the merger of two low-mass helium white dwarfs -- could lead to masses significantly different from the canonical value. The existence of non-canonical massed sdB stars has indeed been suggested, either through asteroseismology \citep{randall06,fontaine19} or from the orbital analysis of close binaries \citep[e.g.][]{ver2022}. In order to illustrate the impact of stellar mass on the EHB location, we also represented tracks computed with STELUM for a core mass of 0.55\,M$_\odot$ in Fig.\,\ref{fig:teff_logg}. These show an important shift toward lower surface gravities and higher effective temperatures where post-EHB objects are usually expected when considering only tracks for the canonical mass. Clearly, this higher-mass EHB can no longer account for the bulk of observed sdB stars, indicating that massive hot subdwarfs must be quite rare. However, it also shows that one cannot reliably assign an evolutionary status to a given sdB star based on its position in the \logg\,--\,\teff\ plane, unless its mass is constrained through independent means. Hence, among the pulsators that are clearly above the 0.47\,M$_\odot$ TAEHB in Fig.\,\ref{fig:teff_logg}, some are likely post-EHB stars, but a few may instead be core helium burning (EHB) objects with a higher mass.

The last information provided in Fig.\,\ref{fig:teff_logg} is the theoretical mapping of the p-mode instability region as predicted by the original calculations of \citet{charpinet01}. This mapping of the \logg\,--\,\teff\ plane results from non-adiabatic pulsation calculations applied to envelope models that incorporate the same nonuniform iron profiles used in STELUM structures. It is well established that the $\kappa$-mechanism involving predominantly iron partial ionization in the $Z$-bump region and enhanced by radiative levitation is responsible for the driving of acoustic pulsations in V361 Hya stars \citep{charpinet96,charpinet97}. The predicted number of radial ($\ell$\,=\,0) p-modes is used in this context as a measure of the driving efficiency. Fig.~\ref{fig:teff_logg} shows that p-mode sdB pulsators clearly concentrate within the region where the driving is predicted to be the strongest (roughly within the three innermost contours), as one would expect, thus supporting that the identification of the main processes which drive oscillations in these stars is correct. The true extent of the p-mode instability region remains uncertain, however. Predicted instabilities (all contours) derived from the assumption of equilibrium between radiative levitation and gravitational settling cover a wider region than the observed instability strip. \citet{charpinet09} demonstrated that the strength of the driving engine is predominantly determined by the amount of heavy metals (in particular iron) present at the $Z$-bump location, irrelevant of the composition in other parts of the envelope. The extent of the instability region is therefore sensitive to the precise amount of heavy metals that can accumulate in that region of the stellar envelope, which is likely modulated by various competing mixing processes that were not all included in the aforementioned calculations (e.g., stellar winds, thermohaline mixing; \citealt{theado09,hu11}). The fact that the observed instability strip is narrower than the predicted one indeed pleads for competing processes that slow down or reduce the efficiency of radiative levitation, the diffusive equilibrium assumption providing an upper limit for the amount of levitating heavy metals. It is also possible that other opaque elements such as nickel play a role in shaping the instability region, although \citet{jeffery06a,jeffery06b,jeffery07} showed that nickel at least had little impact on the p-mode instability strip location, unlike for the g-modes.

Overall, Fig.\,\ref{fig:teff_logg} reveals that the distribution of p-mode pulsators is compatible with the canonical $\sim$0.47\,M$_\odot$ EHB covering all stages of the core helium burning phase, from the ZAEHB to TAEHB. The eight sdB pulsators located well above the TAEHB could be either in a post-EHB, helium shell burning stage or, as discussed above, higher mass sdB stars. The numbers involved, eight versus 60 objects for the main group (i.e. a fraction of one out of eight stars) suggests that higher mass sdBs likely dominate the sample since respective evolution timescales on the EHB and post-EHB would imply a fraction closer to one out of $\sim$30. Hence, statistically speaking (with all due caution when small numbers are involved), among the eight outliers only two should be post-EHB stars and six hot subdwarfs with significantly higher masses. 

Another interesting observation is the concentration of the coolest pulsators near the TAEHB, or equivalently a complete lack of them in earlier stages, closer to the ZAEHB. Considering the size of the sample, we estimate that this trend is real. It could be naturally explained by the fact that, at effective temperatures around 29\,000 K, the intersection of the region of strongest driving efficiency occurs near the TAEHB and does not cover the earlier stages, contrary to the hotter regions located around 34\,000 K where most V361 Hya stars are found. Interestingly, a similar trend, although less obvious, may exist for the hottest pulsators, above $\sim$34\,000 K. If that was to be confirmed, the argument of the driving efficiency as mapped in Fig.\,\ref{fig:teff_logg} would not hold. A potential culprit could then be the growing influence of stellar winds that become more proficient at higher effective temperatures and that could slow down the microscopic diffusion of elements and the onset of favorable conditions for the pulsations to develop. This being posited, we refrain from speculating further on this issue until the reality of the deficit of hot pulsators close to the ZAEHB is confirmed.

\begin{figure}
\centering
\includegraphics[width=\hsize]{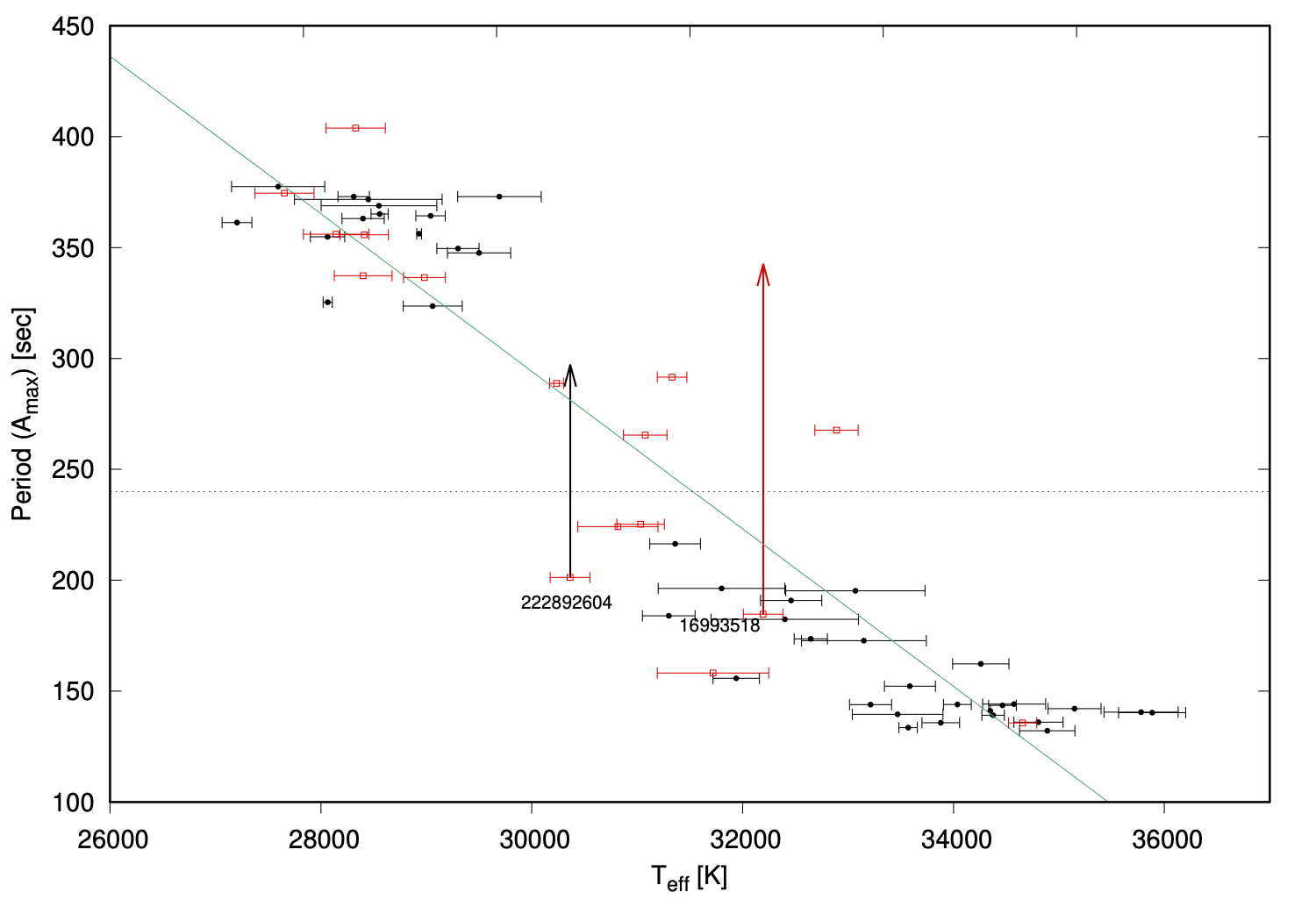}
\caption{Pulsation periods with the highest amplitudes in function of \teff. Black dots represent USC stars, open red boxes represent SC stars, a horizontal dotted line is the Nyquist period, and the solid sloped line is a fit to the USC stars.}
\label{fig:teff_pmax_USC_SC}
\end{figure}
% a*x+b
% a -0.0338 (0.0019)
% b 1309 (62)

\begin{figure}
\centering
\includegraphics[width=\hsize]{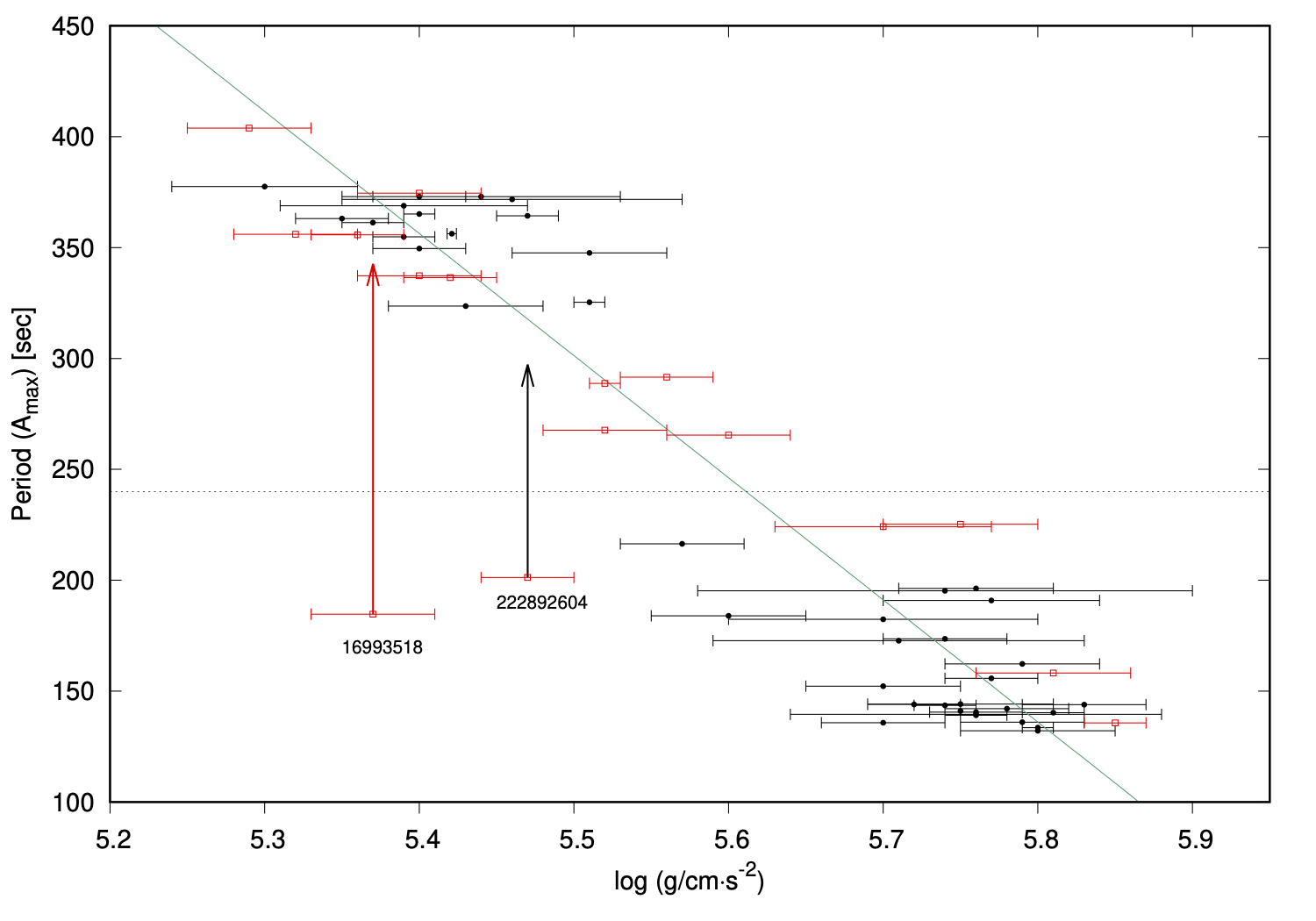}
\caption{Same as in Fig.\,\ref{fig:teff_pmax_USC_SC} but in function of \logg.}
\label{fig:logg_pmax_USC_SC}
\end{figure}
% a*x+b
% a -551 (28)
% b 3330 (159)

\subsection{Pulsation periods vs \teff\ and $log\,g$}
\citet{charpinet02} presented the pulsation period of p-modes as functions of \teff\ and \logg. In their Fig.\,9, they presented results for radial ($\ell$\,=\,0) and non-radial ($\ell$\,=\,1,2,3) modes of radial orders k\,=\,0,1 and 10. Pulsation periods of radial modes do not depend much on \teff\, while they are strongly and monotonically affected by \logg. For non-radial modes both the temperature and gravity can influence the pulsation periods. These change more rapidly with temperature for low radial order cases, and with surface gravity for higher radial order cases. Modal degree does not influence pulsation periods significantly.

Making such an observational plot would have two benefits. Firstly, for a given mode geometry, we can independently derive the effective temperature and/or surface gravity based on a pulsation period. Secondly, if both spectroscopic and period estimates are available, we can test structural models. To compare theoretical and observational counterparts we assume that the frequencies we plot represent the same modal degree and radial order, or that their differences do not significantly change the picture shown in Fig.\,9 of \citet{charpinet02}.

Observational counterpart of pulsation periods in function of \teff\ was reported by e.g. \citet{reed23} who noted an observational correlation between \teff\ and the period of the highest-amplitude pulsation (shown in their Figure\,11, which was a modified version from \citet[][]{reed21}. They examined both p- and g-mode hot subdwarf pulsators, though their sample was heavily dominated by the latter group. Since our work is focused on p-mode pulsators we have significantly updated this figure in the short period range and we also plotted pulsation periods as a function of \logg. We obtained reliable estimates of \teff\ and \logg\ for 41 USC and 16 SC stars, which is 61\% of the p-mode pulsators reported in both Paper I and this paper. This large sample allows us to make a more detailed analysis of the relationship between the period of the highest amplitude (noted as Period(A$_{\rm max}$) in our figures) and \teff/\logg.

As mentioned in Section\,\ref{fourier} we can uniquely identify the frequencies only in the USC data (except for TIC\,220573709). Therefore, we used only USC stars with reliable spectroscopic estimates and plotted pulsation periods as functions of \teff\ and \logg\ in Figs.\,\ref{fig:teff_pmax_USC_SC} and \ref{fig:logg_pmax_USC_SC}. We used pulsation periods with the highest amplitudes, which are not necessarily of the same modal degree and radial order, as presented by \citet{charpinet02} in their Fig.\,9. According to the surface cancellation effect, it is more likely to detect low degree modes, hence we expect those highest amplitude frequencies will likely be assigned with either $\ell$\,=\,0 or 1. The radial order remains unconstrained. Another caveat of our analysis is that the periods having the highest amplitudes can change with time. Depending on which data set is used, a pulsation period shown in Figs.\,\ref{fig:teff_pmax_USC_SC} and \ref{fig:logg_pmax_USC_SC} may not be unique. However, the range of highest-amplitude p-mode pulsation periods in a single star typically does not vary by more than 50\,s for pulsations around 3500\,\uHz\ and decreases toward higher frequencies. Therefore, our selection of periods with the highest amplitudes should not significantly blur the picture in those two figures.

There are noticeable trends in both diagrams, which look more or less linear. However there is significant scatter so it may also be slightly curved/wiggled. Given the scatter in our diagrams we are not able to discern between linear and nonlinear trends plotted in Fig.\,9 of \citet{charpinet02}. We fitted lines to both data sets. Then we added SC stars to see if our sub-/super-Nyquist choices match the trends in the diagrams. The majority of SC stars go along the trend, but we found two stars in the \logg\ diagram (Fig.\,\ref{fig:logg_pmax_USC_SC}) which do not match the fit too well. We marked them with their TIC names. Keeping in mind that our arbitrary super-Nyquist choice could not be correct, we added arrows, which indicate shifts from their super- to sub-Nyquist frequencies. The shift brings the periods of both stars significantly closer to the fit for \logg. However, no shift was necessary in the \teff\ figure as these two stars in that latter figure were not outliers. In the case of TIC\,222892604, the shift moved the star closer to the trend, which supports the shift in the \logg\ diagram. The case of TIC\,16993518 is different. The shift in \logg\ moves the star away from the trend, which does not support a necessity for the shift in the \teff\ diagram. We consider the spectroscopic determination of \logg\ for TIC\,16993518 to be reliable and since all its pulsations are within a narrow range, changing its highest-amplitude frequency would have little effect. Therefore, TIC\,16993518 appears to be an outlier of the trend in the \teff\ diagram.

\section{Summary and conclusions}
We presented results of our search for short-period hot subdwarf pulsators observed in Years 2 and 4 of the \tess\ mission. We identified 50 short-period hot-subdwarf pulsators, in addition to 43 found in Years 1 and 3 and reported by \citet{baran23}. We presented the list of prewhitened frequencies and we made an attempt to identify possible multiplets caused by stellar rotation. We selected five candidates with rotation periods between 11 and 46\,d. Our work completed the search for p-mode pulsating hot subdwarfs in \tess\ Sectors\,1\,--\,60, and allowed us to present a completeness study, discuss an evolutionary status and instability strips of our findings, and compare pulsation periods vs effective temperature and surface gravity with theoretical predictions. We found that the percentage of undetected pulsators in the \tess\ mission reaches 25\% near the 15$^{\rm th}$ magnitude. We underlined the importance of a proper treatment of the hydrogen-rich envelope composition (strongly affected by microscopic diffusion processes) when comparing the distribution of hot subdwarfs in the \logg-\teff\ plane with stellar models. We also emphasized that the stellar mass plays a significant role in understanding the instability strip. Based on the width of the $p$-mode instability strip we deduced that competing mixing processes ignored in the non-adiabatic calculations must play a role to reduce the amount of levitating iron in the stellar envelope. We found that the coolest p-mode pulsators tend to cluster around the Terminal Age of the Extreme Horizontal Branch of $\sim 0.47$ $M_\odot$. Finally, we derived pulsation period distributions that agree with the predicted trends in \teff\ and \logg.

\vfill

\onecolumn % longtable does not work in two-column mode
\twocolumn

\begin{acknowledgements}
Financial support from the National Science Centre Poland under project No.\,UMO-2017/26/E/ST9/00703 is acknowledged. SC acknowledges financial support from the Centre National d’\'Etudes Spatiales (CNES, France) and from the Agence Nationale de la Recherche (ANR, France) under grant ANR-17-CE31-0018. PN acknowledges support from the Grant Agency of the Czech Republic (GA\v{C}R 22-34467S). The Astronomical Institute in Ond\v{r}ejov is supported by the project RVO:67985815. This paper includes data collected with the \tess\ mission, obtained from the MAST data archive at the Space Telescope Science Institute (STScI). Funding for the \tess\ mission is provided by the NASA Explorer Program. STScI is operated by the Association of Universities for Research in Astronomy, Inc., under NASA contract NAS 5–26555. This paper uses observations made with the Nordic Optical Telescope, owned in collaboration by the University of Turku and Aarhus University, and operated jointly by Aarhus University, the University of Turku and the University of Oslo, representing Denmark, Finland and Norway, the University of Iceland
and Stockholm University at the Observatorio del Roque de los Muchachos, La Palma, Spain, of the Instituto de Astrofisica de Canarias. This research has made use of the SIMBAD database, operated at CDS, Strasbourg, France. This paper uses observations from the 2.2m and 3.5m Calar Alto, 2.3m and 4m Kitt Peak, INT (ING), Keck, LAMOST, NTT and VLT (ESO), SOAR, and SDSS telescopes. This work has also made use of data from the European Space Agency (ESA) mission \gaia\ (\url{https://www.cosmos.esa.int/gaia}), processed by the \gaia\ Data Processing and Analysis Consortium (DPAC, \url{https://www.cosmos.esa.int/web/gaia/dpac/consortium}). Funding for the DPAC has been provided by national institutions, in particular the institutions participating in the \gaia\ Multilateral Agreement. V.V.G. is a F.R.S.-FNRS Research Associate. This research has used the services of \url{www.Astroserver.org}.
\end{acknowledgements}

%\begin{Contributions}
%...........
%\end{Contributions}
\vfill

\bibliographystyle{aa}
\bibliography{myrefs}
%\vfill
%\newpage

\begin{appendix}
\section{Additional figures}
\begin{figure*}
\centering
\includegraphics[width=\hsize]{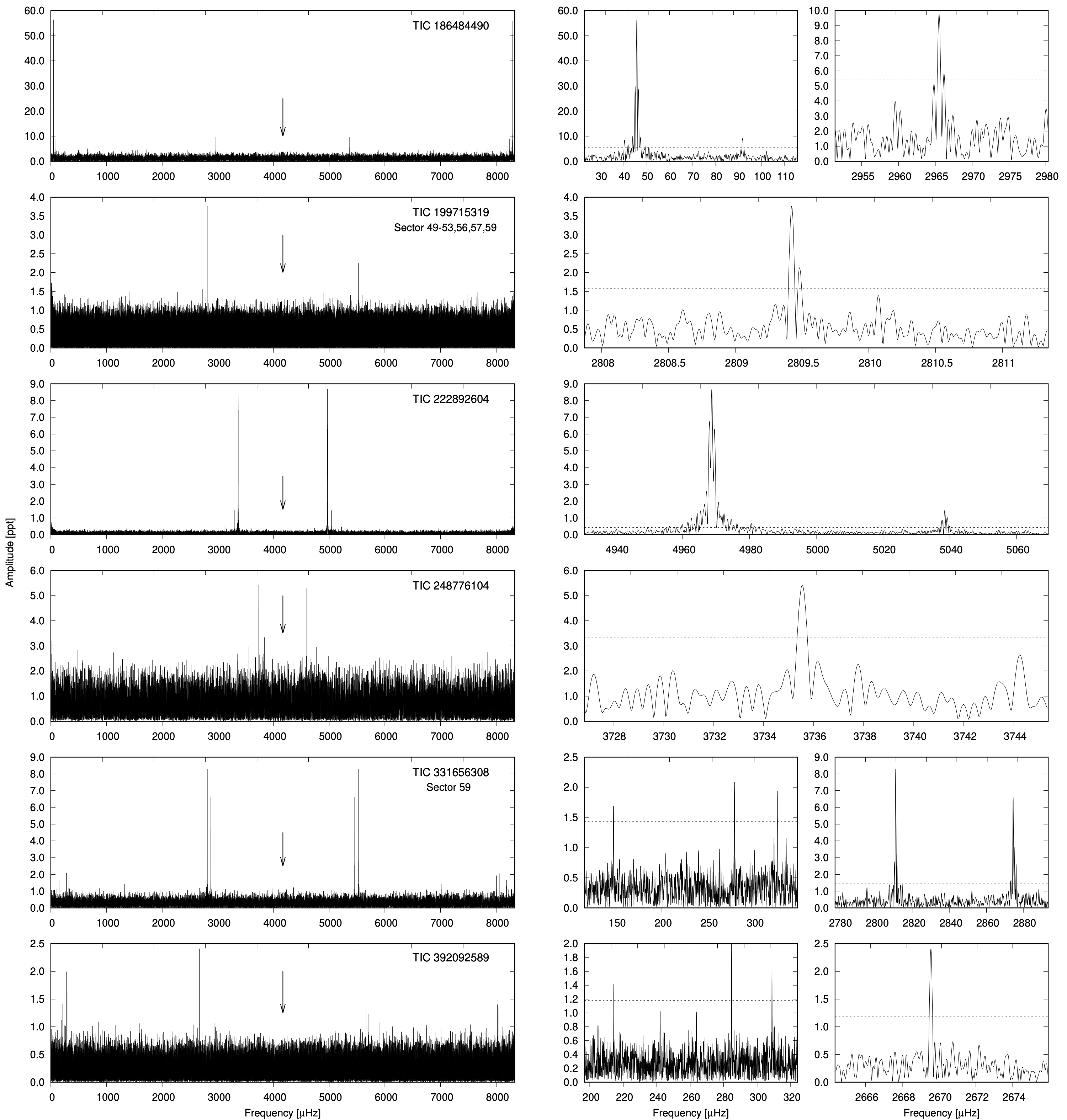}
\caption{Same as in Fig.\,\ref{fig:SC_ft1}, but for another six targets observed only in the SC mode.}
\label{fig:SC_ft2}
\end{figure*}

\begin{figure*}
\centering
\includegraphics[width=\hsize]{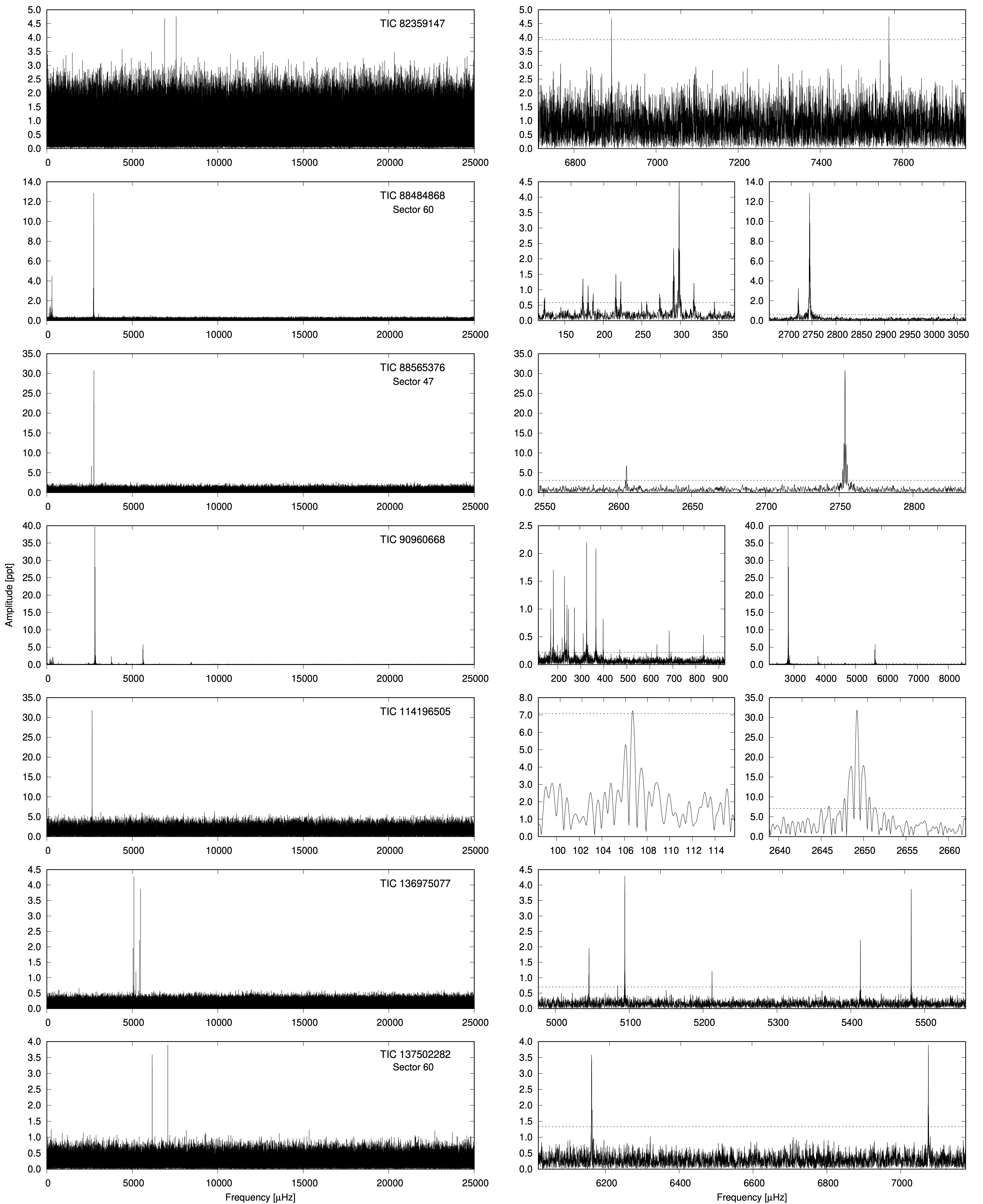}
\caption{Same as in Fig.\,\ref{fig:USC_ft1}, but for another seven targets observed in the USC mode.}
\label{fig:USC_ft2}
\end{figure*}

\begin{figure*}
\centering
\includegraphics[width=\hsize]{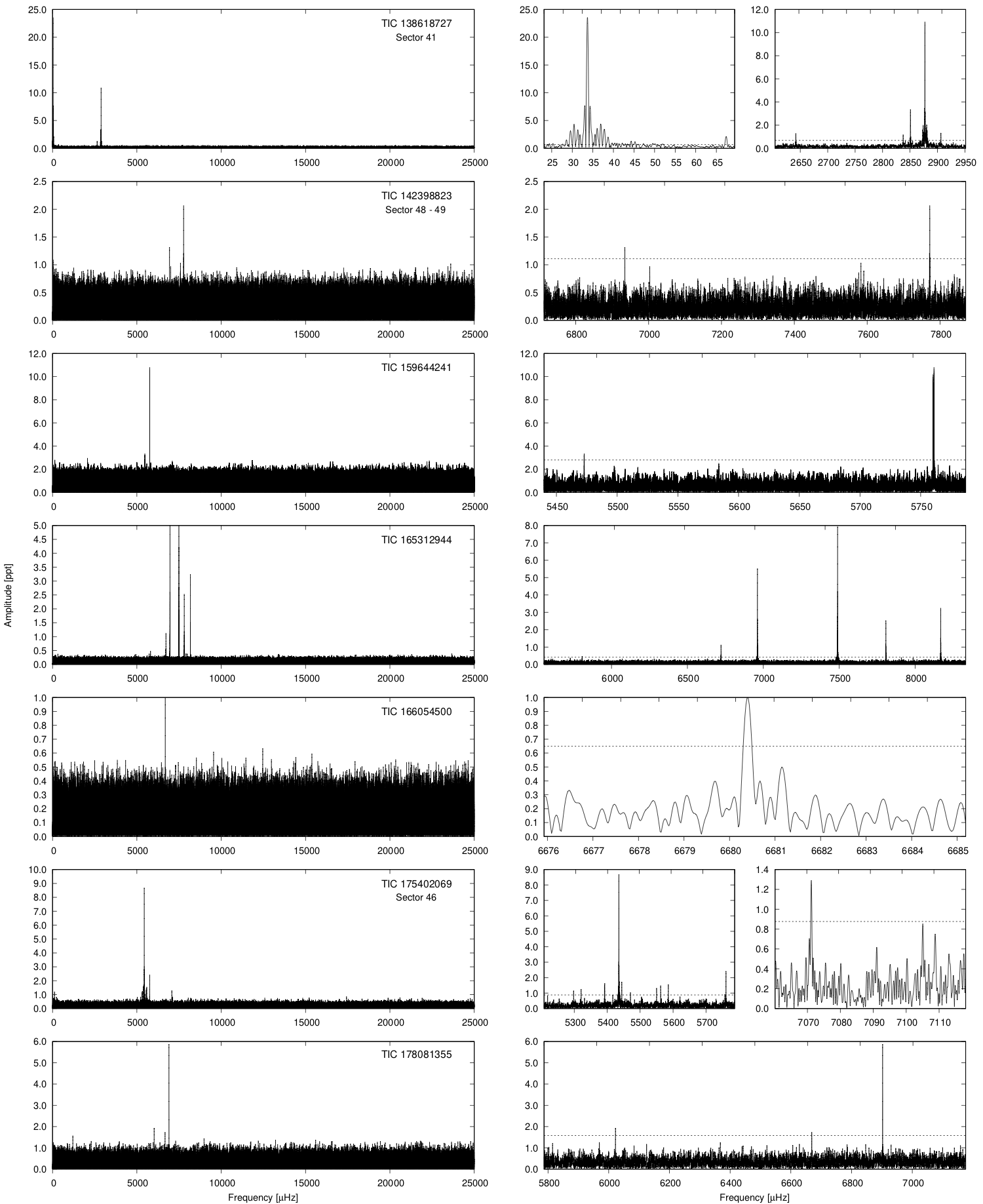}
\caption{Same as in Fig.\,\ref{fig:USC_ft1} but for another seven targets observed in the USC.}
\label{fig:USC_ft3}
\end{figure*}

\begin{figure*}
\centering
\includegraphics[width=\hsize]{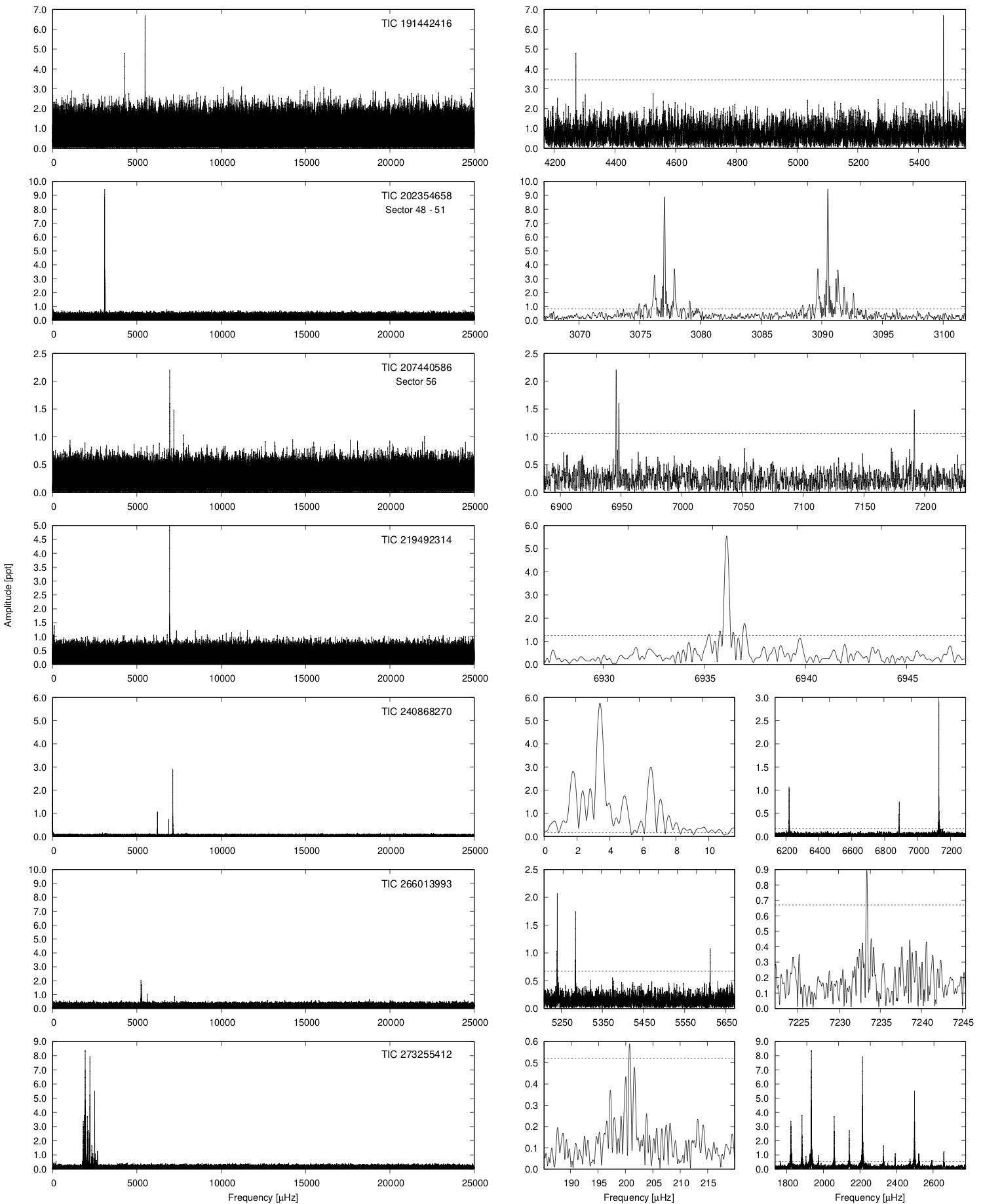}
\caption{Same as in Fig.\,\ref{fig:USC_ft1}, but for another seven targets observed in the USC mode.}
\label{fig:USC_ft4}
\end{figure*}

\begin{figure*}
\centering
\includegraphics[width=\hsize]{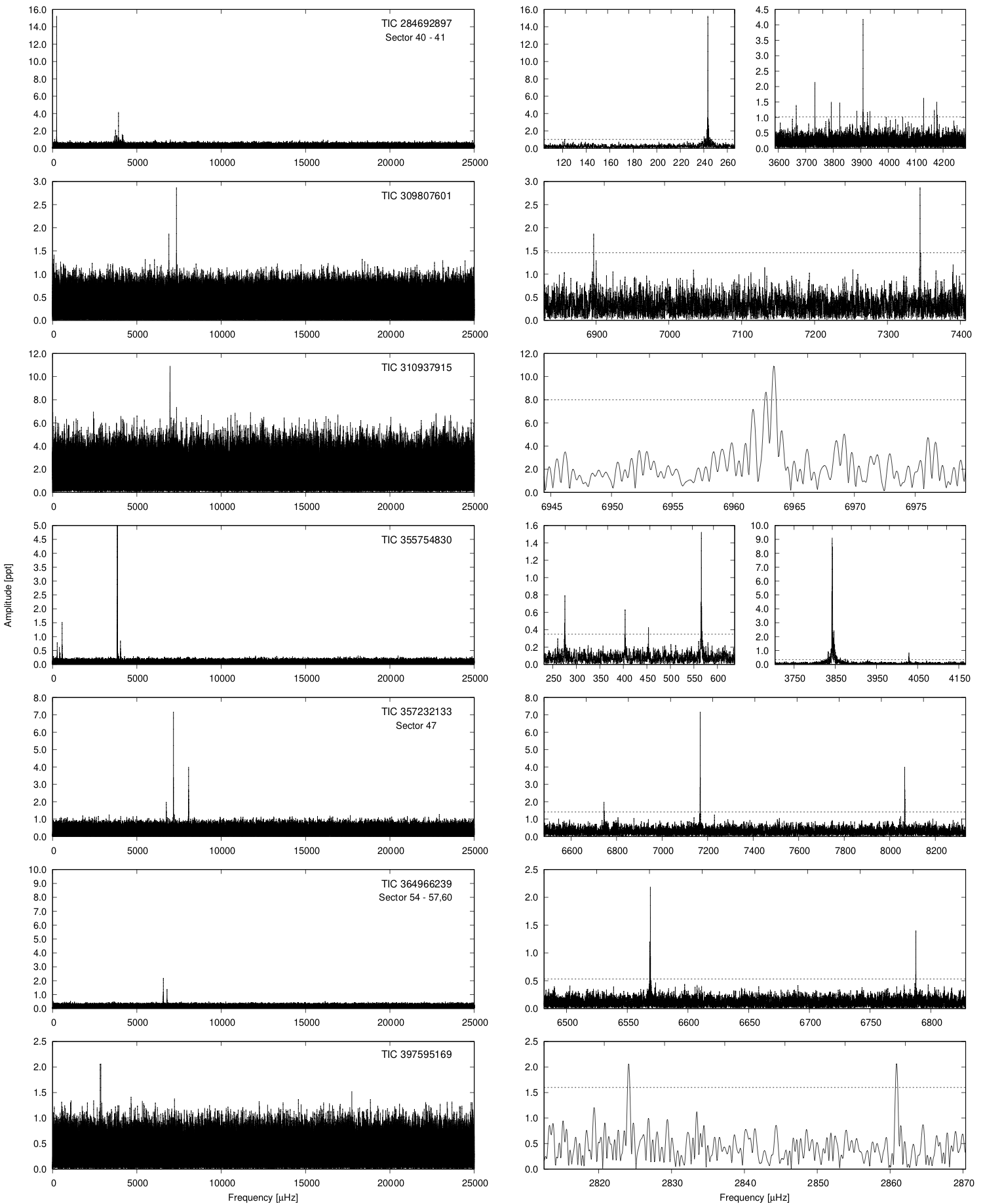}
\caption{Same as in Fig.\,\ref{fig:USC_ft1}, but for another seven targets observed in the USC mode.}
\label{fig:USC_ft5}
\end{figure*}

\begin{figure*}
\centering
\includegraphics[width=\hsize]{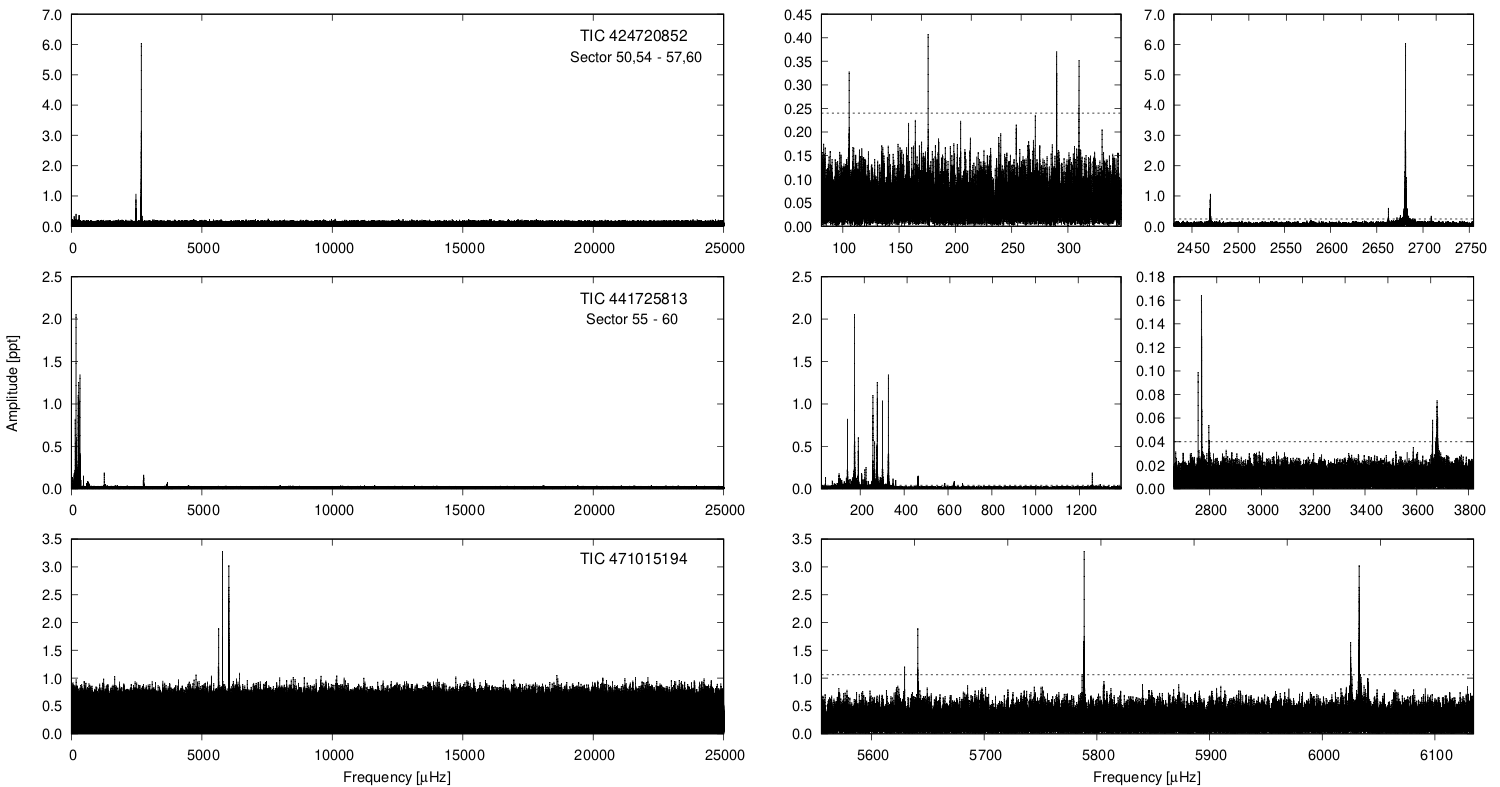}
\caption{Same as in Fig.\,\ref{fig:USC_ft1}, but for another three targets observed in the USC mode.}
\label{fig:USC_ft6}
\end{figure*}

\end{appendix}

\end{document}